\documentclass[jcp,aip,amsmath,amssymb,preprint]{revtex4-2}
\usepackage{float}
\makeatletter
\let\newfloat\newfloat@ltx
\makeatother
\usepackage{graphicx} 
\usepackage[caption=false]{subfig} 
\usepackage{braket}
\usepackage{amsmath}
\usepackage{algorithm}
\usepackage{algpseudocode}
\usepackage{longtable}
\usepackage{multirow}
\usepackage{array}
\usepackage{tikz}
\usepackage{xparse}
\usepackage[colorlinks=true,linkcolor=blue,citecolor=blue,urlcolor=blue]{hyperref}
\usepackage{orcidlink}
\usepackage{mathtools}
\usepackage{parskip}

\usetikzlibrary{arrows.meta,
                positioning,
                calc,
                shapes.geometric,
                math}

\DeclarePairedDelimiter{\norm}{\lVert}{\rVert}

\newcommand{\best}[1]{\textbf{#1}}
\def\equationautorefname~#1\null{Eq.~(#1)\null}
\def\subfigureautorefname~#1\null{Fig.~(#1)\null}
\def\sectionautorefname~#1\null{Sec.~(#1)\null}
\makeatletter
\newenvironment{AlgorithmNoVadjust}{%
  \begin{algorithm}%
  \begingroup
  \let\ALG@old@vadjust\vadjust
  \def\vadjust##1{\relax}%
}{%
  \let\vadjust\ALG@old@vadjust
  \endgroup
  \end{algorithm}%
}
\makeatother

\begin{document}

\title{Tree Tensor Networks Methods for Efficient Calculation of Molecular Vibrational Spectra}

\author{Shuo Sun~\orcidlink{0009-0006-5775-9730}}
\email{shuo.sun@tum.de}
\affiliation{Technical University of Munich, CIT, Department of Computer Science, Boltzmannstra{\ss}e 3, 85748 Garching, Germany}
\author{Richard M.~Milbradt~\orcidlink{0000-0001-8630-9356}}
\affiliation{Technical University of Munich, CIT, Department of Computer Science, Boltzmannstra{\ss}e 3, 85748 Garching, Germany}
\author{Stefan Knecht~\orcidlink{0000-0001-9818-2372}}
\affiliation{Algorithmiq Ltd, Kanavakatu 3C, FI-00160 Helsinki, Finland}
\author{Chandan Kumar~\orcidlink{0000-0001-6510-4204}}
\affiliation{BMW Group, Munich}
\author{Christian B.~Mendl~\orcidlink{0000-0002-6386-0230}}
\email{christian.mendl@tum.de}
\affiliation{Technical University of Munich, CIT, Department of Computer Science, Boltzmannstra{\ss}e 3, 85748 Garching, Germany}
\affiliation{Technical University of Munich, Institute for Advanced Study, Lichtenbergstra{\ss}e 2a, 85748 Garching, Germany}
\date{\today}

\begin{abstract}
We develop and employ general Tree Tensor Networks (TTNs) to compute the vibrational spectra for two model systems: a set of 64-dimensional coupled oscillators and acetonitrile. We explore various tree architectures, ranging from the simple linear structure of Matrix Product States (MPS), to trees where only the leaf nodes carry a physical leg---as commonly seen in the underlying ansatz of the Multilayer Multiconfiguration Time-Dependent Hartree (ML-MCTDH) method---and further to more general trees in which all nodes are allowed to possess a physical leg. In addition, we implement Locally Optimal Block Preconditioned Conjugate Gradient (LOBPCG) methods and Inverse Iteration methods as eigensolvers. Benchmarking runtime and accuracy shows that all tested topologies can reach high accuracy. For acetonitrile, inverse-iteration refinement brings all 84 computed states below 1~cm$^{-1}$ error, while the fork-4 tree, a comb-like tree with four backbone nodes, provides the best overall balance between accuracy and cost. MPS remains computationally attractive, whereas more connected trees generally improve accuracy at fixed bond dimension. All numerical simulations were performed using PyTreeNet, a Python package designed for flexible tensor network computations.
\end{abstract}

\maketitle

\section{Introduction}
\label{sec:introduction}
Vibrational spectroscopies, such as IR and Raman spectroscopies, are widely used in chemistry to characterize molecules. However, interpreting such spectra is not always straightforward, as the vibrational degrees of freedom scale as $3N-6$ (where $N$ is the number of nuclei), making numerical simulations crucial for structure and property determination. In recent years, numerous methods have been developed to obtain vibrational spectra, such as VSCF~\cite{bowman1986self, chaban1999ab, yagi2000direct, roy2013vibrational}, VCI~\cite{rauhut2007configuration, begue2007comparison, neff2009toward, mathea2022advances}, VCC~\cite{christiansen2004vibrational, seidler2009automatic, seidler2011vibrational, thomsen2014optimized}, neural network methods~\cite{Zhang2024}, and quantum simulation~\cite{ollitrault2020hardware, malpathak2025trotter, Majland2025}. VSCF is a mean-field theory and often lacks the accuracy required for the desired applications. More accurate wavefunction-based methods such as VCI and VCC can become difficult to apply as the system size and Hamiltonian complexity increase. Neural network approaches offer a conceptually different framework, but their performance depends strongly on the chosen architecture and training strategy. Quantum simulation methods also incur high costs when simulated classically or require quantum hardware beyond current near-term device capabilities.

As a promising addition to the available methods, tensor network methods have made significant progress in vibrational spectra calculations in recent years~\cite{Wang2014,Rakhuba2016,Baiardi2017,Larsson2019, baiardi2019optimization, Baiardi2021,Larsson2022,glaser2023flexible, glaser2024vibrational, larsson20242500, larsson2024tensor, larsson2025benchmarking, Hoppe2024,Rano2025}. However, most implementations have used matrix product operators (MPOs) to represent the Hamiltonian. This approach limits the full potential of tensor network methods, as the molecules' normal modes exhibit long-range interactions, while the MPO formalism primarily favors nearest-neighbor interactions. This structural limitation is particularly problematic when focusing on higher-lying excited eigenstates, where energy degeneracy becomes significant, making accurate eigenstate computation even more challenging. To this end, we propose in this paper an approach that utilizes tree tensor network operators (TTNOs)~\cite{Frowis2010, Milbradt2024} to represent the Hamiltonian and likewise tree tensor network states (TTNSs)~\cite {Shi2006, gunst2018t3ns} to represent the state function for eigenvalue problem calculations, thereby fully leveraging tree structure connectivity. We implement the Locally Optimal Block Preconditioned Conjugate Gradient (LOBPCG) method and inverse iteration method within the tree tensor network framework to solve the eigenvalue problem. Using a 64-dimensional coupled-oscillator model and acetonitrile~\cite{Carbonniere2004, Leclerc2014} as benchmarks, we compare runtime and accuracy across different tree structures (MPS, T-tree, leaf-only trees, fork trees, and the full T3NS topology; see Appendix~\ref{sec:appendix_tree_structure} and \autoref{fig:ttn_structures_ch3cn_1}, \autoref{fig:ttn_structures_ch3cn_2}). The calculations show that all topologies are viable, that increased connectivity can substantially improve accuracy, and that a moderately branched fork-4 tree offers the best accuracy--cost trade-off for acetonitrile. All computations were performed with PyTreeNet~\cite{PyTreeNet}.

\section{Methods}
\label{sec:methods}

In this section, we first briefly introduce tree tensor networks in \autoref{subsec:ttno_operations}. We then discuss two well-known numerical eigenvalue algorithms in \autoref{subsec:numerical algorithms}. Finally, we combine these two aspects and introduce our implementation of these numerical algorithms within the tensor network framework in \autoref{subsec:implementation in ttns}.

\subsection{Tree Tensor Network Operations}
\label{subsec:ttno_operations}

\subsubsection{Introduction to Tensor Networks}

The field of tensor networks has seen significant research activity over the last two decades. It has been especially successful when applied to the numerical simulation of quantum systems. We will restrict this introduction to the concepts relevant to the present work. For a deeper and pedagogical introduction, refer to \cite{Schollwock2011, Evenbly2022, Bridgeman2017, Silvi2019}, and for a general overview, see the recent reviews \cite{Orus2019, Cirac2021, Banuls2023}. In general, a tensor is an element $A \in \mathbb{C}^{\times_{i=1}^N D_i}$, i.e., a multi-dimensional array whose entries are identified by a string of $N$ indices, where the $i$th index can take values in $\{ 1, \dots, D_i\}$. Intuitively, this is a generalization of vectors and matrices, where the entries are identified by one and two indices, respectively. We will call $N$ the \emph{degree} of a tensor, while we refer to the different indices on a tensor as \emph{legs}.\\

If a leg $i$ of a tensor $A$ and a leg $i'$ of a different tensor $B$ have the same dimension $D_i=D_{i'}$, we can contract over these legs by summing over the products of the tensor entries with the same index for the contracted legs. Let us emphasize this with an example, where $A \in \mathbb{C}^{D_1\times D_2 \times D_3}$ and $B \in \mathbb{C}^{D_2 \times D_4 \times D_5}$. Here we can contract the second leg of $A$ with the first leg of $B$, leading to a new tensor $C \in \mathbb{C}^{D_1 \times D_3 \times D_4 \times D_5}$. The entries of $C$ are defined by
\begin{equation}\label{eq:tensor_contr_eq}
    C_{jk\ell m} = \sum_{n=1}^{D_2} A_{jnk} B_{n\ell m}.
\end{equation}
Now, a \emph{tensor network} is simply a collection of tensors and a record of the legs that are contracted. To facilitate easier readability, a graphical notation for tensor networks is available. One can build this representation by depicting every tensor as a geometrical shape and attaching one line per leg. Then, all lines representing contracted legs are connected to their corresponding partners. For example, \autoref{eq:tensor_contr_eq} can be depicted as
\begin{equation}
    \raisebox{-0.5cm}{
    \begin{tikzpicture}
        \node[inner xsep=0.2cm, draw, fill=teal!80, very thick] (C) at (0,0) {$C$};
        \node at (0.2,0.6) {$j$};
        \node at (0.6,0.2) {$\ell$};
        \node at (-0.3,-0.6) {$k$};
        \node at (0.4,-0.6) {$m$};
        \draw (0,1) -- (C);
        \draw (-0.2,-1) -- (C);
        \draw (0.2,-1) -- (C);
        \draw (1,0) -- (C);
        \node at (1.3,0) {$=$};
        \node[circle, draw, fill=cyan!30, very thick] (A) at (2,0) {$A$};
        \node[circle, draw, fill=cyan!50, very thick] (B) at (3.5,0) {$B$};
        \node at (1.8,0.6) {$j$};
        \node at (2.7,0.2) {$n$};
        \node at (1.8,-0.6) {$k$};
        \node at (4.1,0.2) {$\ell$};
        \node at (3.3,-0.6) {$m$};
        \draw (2,1) -- (A) -- (2,-1);
        \draw (A) -- (B);
        \draw (3.5,-1) -- (B) -- (4.5,0);
    \end{tikzpicture}
    }.
\end{equation}

We can also split a tensor. To this end, we can reinterpret a tensor as a matrix by defining which legs should be the output and input legs of the matrix. Then we can apply any desired matrix decomposition, the most common being the QR decomposition and the singular value decomposition (SVD). For example, we can obtain two tensors, say $A'$ and $B'$, from a tensor $C$ by first reshaping $C$ into a matrix. This is done by grouping some of its legs to form the `row' indices and the remaining legs to form the `column' indices. After applying a matrix decomposition (e.g., SVD or QR), the resulting matrices can be reshaped back into tensors, effectively splitting the original tensor $C$ into two or more tensors. These two kinds of tensor manipulation allow us to represent an arbitrary quantum state as a tensor network, as we discuss in the next section.

\subsubsection{Tensor Network States and Operators}
A quantum state $\ket{\psi}$ generally has a large number of degrees of freedom scaling exponentially in system size. For example, if we have a quantum system consisting of $L$ smaller quantum systems, we can write the state as
\begin{equation}
    \ket{\psi} = \sum_{s_1, \dots, s_L} C_{s_1, \dots, s_L} \ket{s_1, \dots , s_L},
\end{equation}
where $\{ \ket{s_j} \}$ is the basis of the local quantum system $j$. Repeated tensor decompositions allow us to express the tensor $C$ as a tensor network with an arbitrary underlying tree shape
\begin{equation}\label{eq:ttns}
\ket{\psi} = \sum_{\substack{s_1, \ldots, s_L\\
v_1, \ldots, v_M}} T^{(1)}_{\{v_{1,i}\}, s_1} \ldots T^{(L)}_{\{v_{L,i}\}, s_L} \ket{s_1, s_2, \ldots, s_L },
\end{equation}
where we assumed for simplicity of notation that every tensor has one uncontracted leg $s_j$. Here $s_j$ indexes the \emph{physical} (local system) degrees of freedom with dimension called the physical leg dimension, while $v_{(j,i)}$ indexes the \emph{virtual} (internal) degrees of freedom connecting tensors in the network. The dimension of $v_i$ is denoted as virtual bond dimension. The total number of contracted legs in the tensor network is $M$ and $\{v_{j,i}\} \subset \{v_1, \dots, v_M\}$ are the contracted legs of tensor $T^{(j)}$. The tensor network in \autoref{eq:ttns} is a \emph{tree tensor network state (TTNS)} if there are no loops in the network\cite{Shi2006}.

We can apply the analogous decomposition to a quantum operator acting on high-dimensional systems, yielding
\begin{equation}\label{eq:ttno}
    \hat{O} = \sum_{\substack{p_1, \ldots, p_L\\
                              q_1, \ldots, q_L\\
                              v_1, \ldots, v_M}} 
                    \mathcal{T}^{(1)}_{\{v_{1,i}\}, p_1, q_1} \cdots \mathcal{T}^{(L)}_{\{v_{L,i}\}, p_L, q_L} \ket{p_1, p_2, \ldots, p_L}\bra{q_1, q_2, \ldots, q_L}.
\end{equation}
This tensor network is called a \emph{tree tensor network operator (TTNO)} \cite{Frowis2010} and differs from a TTNS mostly in that every site has two uncontracted physical legs rather than just one. A graphical depiction of an example TTNS and TTNO based on the same tree is shown in \autoref{fig:trees}.

\begin{figure*}[ht]
    \centering
    \subfloat[TTNS]{
        \centering
        \begin{tikzpicture}[state/.style={circle, draw, fill=cyan!70, ultra thick, minimum size=6mm, inner sep=0pt},
  operator/.style={regular polygon,regular polygon sides=4, draw, fill=teal!30, ultra thick, minimum size=8mm, inner sep=0pt},]
            \def\vertdist{1}
            \def\horizdist{1}
            \def\physdist{0.75}
            
            \node[state,draw] (N1) at (0,0) {$T^{(1)}$};
            \node[state, draw] (N2) at (-\horizdist,-\vertdist) {$T^{(2)}$};
            \node[state, draw] (N3) at (\horizdist,-\vertdist) {$T^{(3)}$};
            \node[state, draw] (N4) at (0*\horizdist,-2*\vertdist) {$T^{(4)}$};
            \node[state, draw] (N5) at (2*\horizdist,-2*\vertdist) {$T^{(5)}$};
            
            \draw[thick] (N2) -- (N1) -- (N3) -- (N4);
            \draw[thick] (N3) -- (N5);
            
            \draw[very thick] (N1) -- (0,\physdist);
            \draw[very thick] (N2) -- (-\horizdist,-\vertdist+\physdist);
            \draw[very thick] (N3) -- (\horizdist,-\vertdist+\physdist);
            \draw[very thick] (N4) -- (0*\horizdist,-2*\vertdist+\physdist);
            \draw[very thick] (N5) -- (2*\horizdist,-2*\vertdist+\physdist);
        \end{tikzpicture}
        \label{fig:ttns}
    }
    \subfloat[TTNO]{
        \centering
        \begin{tikzpicture}[operator/.style={trapezium, trapezium left angle=70, trapezium right angle=110, minimum width=5mm, minimum height=5mm, draw, fill=teal!30, ultra thick, inner sep=0pt}]
            \def\vertdist{1}
            \def\horizdist{1}
            \def\physdist{0.75}
            
            \node[operator,draw] (N1) at (0,0) {$\mathcal{T}^{(1)}$};
            \node[operator, draw] (N2) at (-\horizdist,-\vertdist) {$\mathcal{T}^{(2)}$};
            \node[operator, draw] (N3) at (\horizdist,-\vertdist) {$\mathcal{T}^{(3)}$};
            \node[operator, draw] (N4) at (0*\horizdist,-2*\vertdist) {$\mathcal{T}^{(4)}$};
            \node[operator, draw] (N5) at (2*\horizdist,-2*\vertdist) {$\mathcal{T}^{(5)}$};
            
            \draw[thick] (N2) -- (N1) -- (N3) -- (N4);
            \draw[thick] (N3) -- (N5);
            
            \draw[very thick] (N1) -- (0,\physdist);
            \draw[very thick] (N2) -- (-\horizdist,-\vertdist+\physdist);
            \draw[very thick] (N3) -- (\horizdist,-\vertdist+\physdist);
            \draw[very thick] (N4) -- (0*\horizdist,-2*\vertdist+\physdist);
            \draw[very thick] (N5) -- (2*\horizdist,-2*\vertdist+\physdist);
            \draw[very thick] (N1) -- (0,-\physdist);
            \draw[very thick] (N2) -- (-\horizdist,-\vertdist-\physdist);
            \draw[very thick] (N3) -- (\horizdist,-\vertdist-\physdist);
            \draw[very thick] (N4) -- (0*\horizdist,-2*\vertdist-\physdist);
            \draw[very thick] (N5) -- (2*\horizdist,-2*\vertdist-\physdist);
        \end{tikzpicture}
        \label{fig:ttno}
    }
    \caption{Examples of a TTNS and TTNO with the same tree topology, differing in the number of physical legs per tensor. Adapted from Çakır \textit{et al.}, Phys. Rev. B, 112, 035101, 2025; licensed under a Creative Commons Attribution (CC BY) license.}
    \label{fig:trees}
\end{figure*}

\subsubsection{TTNO Construction}
\label{subsubsec:ttno_construction}

To construct a TTNO representing a desired system Hamiltonian, we use the algorithm introduced in Ref.~\onlinecite{Cakir2025}, which improves upon the methods introduced in Refs.~\onlinecite{Milbradt2024, li2024optimal}. The algorithm starts from a Hamiltonian in the form
\begin{equation}\label{eq:sum_of_tp}
    H = \sum_{i=1}^N c_i \bigotimes_{j=1}^L A_i^{(j)},
\end{equation}
where $A_i^{(j)}$ is an operator acting on the $j$-th subsystem (or site). We denote the overall number of subsystems by $L$. Once we have decided on a desired tree structure, we rewrite the Hamiltonian as a \emph{state diagram}. State diagrams are a symbolic, hypergraph-based representation of tensor network topologies. Here, each path through the graph corresponds to a term in the Hamiltonian sum. This intermediate representation allows us to exploit the tensor-product structure of the Hamiltonian to systematically combine repeated operators. To be more specific, the hyperedges $h_{i,j}$, where $j$ again denotes the subsystem, in the state diagram are labelled and correspond to the operator-valued elements of the TTNO tensors. On the other hand, the vertices $v_{\ell,(j,j')}$ connected by the hyperedges each correspond to one index value $\ell$ of the virtual leg connecting $j$ and $j'$ in the TTNO~\cite{Milbradt2024}.

It is trivial to initialise the state diagram for all terms individually. Combined into one big state diagram representing the Hamiltonian, they will be unconnected, though they contain superfluous information. Thus, we compress the state diagrams by combining equivalent parts of the different single-term state diagrams. However, this is a non-trivial task. To approach this, we traverse the underlying tree structure and, for each edge, compress the corresponding set of vertices $\mathcal{V}$ in the state diagram. The compression occurs in multiple phases: First, the state diagram is split into two parts $\mathcal{U}_1$ and $\mathcal{U}_2$ by removing $\mathcal{V}$. Now, any equivalent elements in each part $\mathcal{U}_i$ will be represented by one element. We can obtain a labelled bipartite graph by reinstating the connectivity between $\mathcal{U}_1$ and $\mathcal{U}_2$ due to $V$; the labels correspond to the prefactors $c_i$ of the terms. We can find the optimal placement of new vertices via an application of a symbolic version of the Gaussian elimination algorithm on the label matrix of the bipartite graph~\cite{Cakir2025}, followed by the minimum vertex cover algorithm on the result~\cite{li2024optimal}.

Once we compress the entire state diagram, it can be translated into a TTNO. To do so, one runs over all the labelled hyperedges $h_{i,j}$. The corresponding operator is then added to the tensor $\mathcal{T}^{[j]}$ at the indices denoted by the vertices connected to $h_{i,j}$~\cite{Milbradt2024}.

\subsection{Numerical Algorithms}
\label{subsec:numerical algorithms}

A variety of numerical approaches have been developed for computing vibrational excited states, including improved relaxation~\cite{Wang2014}, LOBPCG and inverse iteration~\cite{Rakhuba2016}, DMRG~\cite{Baiardi2017, Larsson2019, Larsson2022}, FEAST~\cite{Baiardi2021}, inexact Lanczos~\cite{Rano2025}, state-averaged DMRG~\cite{Larsson2022}, state-averaged ML-MCTDH-based Lanczos~\cite{Hoppe2024}, as well as shift-and-invert, folded DMRG, and excited-state-specific DMRG~\cite{baiardi2019optimization}. 

In this work, we adapt and employ the LOBPCG and inverse iteration eigensolvers of Ref.~\onlinecite{Rakhuba2016} within the TTN framework. Both methods rely on several TTN-specific operations, which are detailed in \autoref{subsec:implementation in ttns}. These include the evaluation of $H\ket{x}$, compression of sums of states using successive randomized compression (SRC,~\autoref{subsubsec:src}), and the solution of linear systems of the form $H\ket{y} = \ket{x}$ via alternating least squares (ALS,~\autoref{subsubsec:als}). In addition, LOBPCG requires the recompression of linear combinations of TTNSs, which can be carried out using either variational fitting (\autoref{subsubsec:variational_fitting}) or SRC. 

In the following, we first outline the general form of the eigensolvers and then describe their specialization to TTNs.

\subsubsection{Locally Optimal Block Preconditioned Conjugate Gradient (LOBPCG) method}
\label{subsubsec:lobpcg}
Locally Optimal Block Preconditioned Conjugate Gradient (LOBPCG) is a numerical method to solve eigenvalue problems~\cite{knyazev2001, stathopoulos2002}. It is a generalization of the conjugate gradient method, which is very efficient for large sparse matrices. In this work, we adapt LOBPCG to tree tensor networks. The core idea is to construct, at each iteration, a search subspace spanned by the current approximate eigenvector, its residual, and the previous search direction, and then solve a Rayleigh--Ritz problem within this subspace.

We begin by initializing the residual vector as $r = H x - \lambda x$, where $x$ is the current state vector and $\lambda$ is the Rayleigh quotient,
\begin{align}
    \lambda = \frac{x^\dagger H x}{x^\dagger x}.
\end{align}
In this work we use the three-term recurrence formulation of LOBPCG,
\begin{subequations}
\label{eq:lobpcg update}
\begin{align}
    p^{(k)} &= \beta r^{(k)} + \gamma p^{(k-1)} \\
    \tilde{x}^{(k+1)} &= \alpha x^{(k)} + p^{(k)} \\
    x^{(k+1)} &= \frac{\tilde{x}^{(k+1)}}{\|\tilde{x}^{(k+1)}\|}
\end{align}
\end{subequations}
where the coefficients $\alpha$, $\beta$, and $\gamma$ are obtained by solving the small generalized eigenvalue problem
\begin{equation}   
    \begin{bmatrix}
    \begin{pmatrix}
        x^{(k)} \\
        r^{(k)} \\
        p^{(k)}
    \end{pmatrix}
    H 
    \begin{pmatrix}
        x^{(k)} \quad r^{(k)} \quad p^{(k)}
    \end{pmatrix}
    \end{bmatrix}
    \begin{pmatrix}
        \alpha \\
        \beta \\
        \gamma
    \end{pmatrix}
    = \lambda
    \begin{bmatrix}
    \begin{pmatrix}
        x^{(k)} \\
        r^{(k)} \\
        p^{(k)}
    \end{pmatrix}
    \begin{pmatrix}
        x^{(k)} \quad r^{(k)} \quad p^{(k)}
    \end{pmatrix}
    \end{bmatrix}
    \begin{pmatrix}
        \alpha \\
        \beta \\
        \gamma
    \end{pmatrix}.
\end{equation}

This method can also be implemented in a block manner, where we compute $N$ eigenvalues and eigenvectors simultaneously by constructing $x^{(k)}$, $r^{(k)}$, and $p^{(k)}$ for each eigenvector and solving a $3N \times 3N$ generalized eigenvalue problem. This block formulation is particularly important for handling near-degenerate states (as mentioned in \autoref{sec:introduction}), as it solves for multiple eigenstates within the same subspace, preventing convergence to the same state.

To accelerate convergence, we can apply a preconditioner $M$ to the residual vector, which is typically the inverse of the shifted Hamiltonian, 
\begin{align}
	r_{\text{precond}} & = M r = (H-s\mathcal{I})^{-1} r,\label{eq:precondition}
\end{align}
where $s$ is the shift value and $\mathcal{I}$ is the identity matrix. The preconditioned residual can be obtained by solving the linear system 
\begin{align}
(H-s\mathcal{I}) r_\text{precond} = r.\label{eq:solve precondition}
\end{align}
In the tree tensor network framework, we solve this equation by iterating through the network nodes (called sweeps), optimizing one tensor at a time, as described in~\autoref{subsubsec:als}.

The pseudocode of the single and block LOBPCG method is shown in \autoref{alg:lobpcg_single} and \autoref{alg:lobpcg_block}, respectively. Note that at the start of the first iteration, the search directions are initialized as $p^{(0)} = r^{(0)}$. For all subsequent iterations, $p^{(i)}$ is updated from the generalized eigenvector components according to~\autoref{eq:lobpcg update}.

\begin{AlgorithmNoVadjust}
\caption{LOBPCG for a Single State}
\label{alg:lobpcg_single}
\begin{algorithmic}[1]
\Require Matrix $H$, initial state vector $x^{(0)}$, shift for preconditioner $s$, max iterations $N_{\text{iter}}$
\Ensure Converged state $x$ and eigenvalue $\lambda$
\State Initialize: $x \leftarrow x^{(0)}$, $\lambda \leftarrow \frac{ x^\dagger H x}{x^\dagger x }$
\For{$i = 1$ to $N_{\text{iter}}$}
    \State Compute residual: $r \leftarrow Hx - \lambda x$ \Comment{via successive randomized compression and  variational fitting}
    \State Apply preconditioner: $r \leftarrow M r$ \Comment{See~\autoref{eq:precondition}, via alternating least squares~(ALS)}
    \State Build subspace: $\mathcal{S} \leftarrow \{x, r, p\}$ \Comment{For $i=1$, set $p \leftarrow r$}
    \If{orthogonalization required}
        \State Orthogonalize all vectors in $\mathcal{S}$ against the constraint subspace \Comment{e.g., previously computed lower-lying eigenstates}
    \EndIf
    \State Construct matrices: $\tilde{H}_{kl} \leftarrow s_k^\dagger H s_l$, $\tilde{M}_{kl} \leftarrow  s_k^\dagger s_l$ for $s_k, s_l \in \mathcal{S}$
    \State Solve generalized eigenvalue problem: $\tilde{H}\mathbf{v} = \lambda \tilde{M}\mathbf{v}$, $\mathbf{v}=(\alpha, \beta, \gamma)^T $
    \State Update $x$, $p$ from eigenvectors, $\lambda \leftarrow$ smallest eigenvalue \Comment{See~\autoref{eq:lobpcg update}}
\EndFor
\State \Return $x$, $\lambda$
\end{algorithmic}
\end{AlgorithmNoVadjust}

\begin{AlgorithmNoVadjust}
\caption{Block LOBPCG for Multiple States}
\label{alg:lobpcg_block}
\begin{algorithmic}[1]
\Require Matrix $H$, initial state vectors $\{x_j^{(0)}\}_{j=1}^{n}$, shift for preconditioner $s$, max iterations $N_{\text{iter}}$
\Ensure Multiple eigenstates $\{x_j\}_{j=1}^{n}$ and eigenvalues $\{\lambda_j\}_{j=1}^{n}$
\State Initialize: $\{x_j\}_{j=1}^{n} \leftarrow \{x_j^{(0)}\}_{j=1}^{n}$, $\{\lambda_j\}_{j=1}^{n} \leftarrow \left\{\frac{ x_j^\dagger H x_j}{x_j^\dagger x_j }\right\}_{j=1}^{n}$
\For{$i = 1$ to $N_{\text{iter}}$}
    \For{$j = 1$ to $n$}
        \State Compute residual: $r_j \leftarrow Hx_j - \lambda x_j$ \Comment{via successive randomized compression, variational fitting}
        \State Apply preconditioner: $r_j \leftarrow M r_j$ \Comment{See~\autoref{eq:precondition}, via alternating least squares~(ALS)}
        \If{orthogonalization required}
            \State Orthogonalize $x_j$, $r_j$ and $p_j$ against the constraint subspace
        \EndIf
    \EndFor
    \State Build subspace: $\mathcal{S} \leftarrow \{x_j\}_{j=1}^{n} \cup \{r_j\}_{j=1}^{n}\cup \{p_j\}_{j=1}^{n}$
    \State Construct matrices: $\tilde{H}_{k\ell} \leftarrow  s_k^\dagger H s_{\ell}$, $\tilde{M}_{k\ell} \leftarrow s_k^\dagger s_{\ell}$ for $s_k, s_{\ell} \in \mathcal{S}$
    \State Solve generalized eigenvalue problem: $\tilde{H}\mathbf{v} = \lambda \tilde{M}\mathbf{v}$
    \For{$j = 1$ to $n$}
        \State Update state: $x_j \leftarrow \sum_{k=1}^{n} v_{k,j} s_k$
            \State $p_j \leftarrow \sum_{k=n+1}^{3n} v_{k,j} s_k$ \Comment{See~\autoref{eq:lobpcg update}}
    \EndFor
    \State Update eigenvalues: $\{\lambda_j\}_{j=1}^{n} \leftarrow$ smallest $n$ eigenvalues
\EndFor
\State \Return $\{x_j\}_{j=1}^{n}$, $\{\lambda_j\}_{j=1}^{n}$
\end{algorithmic}
\end{AlgorithmNoVadjust}

\subsubsection{Inverse Iteration method}
\label{subsubsec:inverse_iteration}
Inverse iteration is an iterative method to find the eigenvector and eigenvalue of a matrix close to a given shift value. The basic idea is that 
\begin{subequations}
\label{eq:inverse iteration update}
\begin{align}
    \tilde{x}^{(i+1)} &= (H - s\mathcal{I})^{-1} x^{(i)} \\
    x^{(i+1)} &= \frac{\tilde{x}^{(i+1)}} {\lVert \tilde{x}^{(i+1)} \rVert}
\end{align}
\end{subequations}
where again $s$ is the shift value and $\mathcal{I}$ the identity matrix. When $s$ is not exactly one of the eigenvalues of $H$, we can obtain the updated eigenvector by solving the linear system $(H - s\mathcal{I}) x^{(i+1)} = x^{(i)}$ by sweeping through the tensor network, similar to~\autoref{eq:solve precondition}. In this work, we found the solution of the linear system using the alternating least squares (ALS) method, which is described in~\autoref{subsubsec:als}. The shift value can stay the same throughout the iterations and may also be updated after one or few iterations. In this project, we cluster eigenvalues into groups and use block inverse iteration to solve the eigenproblem. Clustering is particularly important for near-degenerate eigenvalues where individual shifts might miss multiple closely spaced states. We use the mean Rayleigh quotient of the group as the shift value and update the Rayleigh quotient in each iteration. The pseudocode of the block inverse iteration method is described in~\autoref{alg:block_inverse_iteration}.

\begin{AlgorithmNoVadjust}
\caption{Block Inverse Iteration}
\label{alg:block_inverse_iteration}
\begin{algorithmic}[1]
\Require Matrix $H$, initial state vectors $\{x_j^{(0)}\}_{j=1}^{n}$, max iterations $N_{\text{iter}}$
\Ensure Multiple eigenstates $\{x_j\}_{j=1}^{n}$ and eigenvalues $\{\lambda_j\}_{j=1}^{n}$
\State Initialize: $\{x_j\}_{j=1}^{n} \leftarrow \{x_j^{(0)}\}_{j=1}^{n}$, $\{\lambda_j\}_{j=1}^{n} \leftarrow \left\{\frac{ x_j^\dagger H x_j}{x_j^\dagger x_j }\right\}_{j=1}^{n}$
\For{$i = 1$ to $N_{\mathrm{iter}}$}
    \State Update shift and Hamiltonian: $s \leftarrow \frac{1}{n}\sum_{j=1}^{n} \lambda_j$, $H_\mathrm{shift} \leftarrow H - s \mathcal{I}$ 
    \For{$j = 1$ to $n$}
        \State Solve $H_{\mathrm{shift}}y_j = x_j$, update state: $x_j \leftarrow y_j/\norm{y_j}$ \Comment{Solve with ALS~\autoref{subsubsec:als}}
    \EndFor
    \If{$n > 1$}
        \State Orthogonalize vectors by solving generalized eigenvalue problem \State $\tilde{H}_{k\ell} \leftarrow x_k^\dagger H x_{\ell}, \tilde{M}_{k\ell} = x_k^\dagger x_{\ell}$, $\tilde{H} \mathrm{v} = \lambda \tilde{M} \mathbf{v}$, $x_j \leftarrow \sum_{k} v_{jk} x_k$
    \EndIf  
\EndFor
\State \Return $\{x_j\}_{j=1}^{n}$, $\{\lambda_j\}_{j=1}^{n}$
\end{algorithmic}
\end{AlgorithmNoVadjust}

\subsection{Implementation of LOBPCG and Inverse Iteration on Tree Tensor Networks}
\label{subsec:implementation in ttns}

\subsubsection{Successive Randomized Compression}
\label{subsubsec:src}
The successive randomized compression (SRC) method~\cite{Camao2026} is used to approximate the product of a TTNO $H$ and a TTNS $x$. It was originally developed for linear tensor networks, and we extend it here to arbitrary tree structures~\cite{Milbradt2025}. The naive approach to applying a TTNO to a TTNS is to contract the local operator and state tensors explicitly and then compress the resulting network by an SVD sweep. This procedure scales poorly with the TTNS and TTNO bond dimensions (see Appendix~\ref{sec: scaling}). SRC avoids the most expensive parts of that direct contraction.

SRC~\cite{Camao2026} draws on ideas from randomized linear algebra~\cite{Halko2011, Murray2023}. For a large matrix $A \in \mathbb{C}^{M\times N}$, one approximates it as
\begin{equation}
\tilde{A} = QB,\label{eq:matrix random compression}
\end{equation}
where $Q\in \mathbb{C}^{M \times p}$ has orthonormal columns and $B\in\mathbb{C}^{p \times N}$. This is achieved by drawing a random matrix $\Omega\in\mathbb{C}^{N\times p}$, forming the sketch $Y=A\Omega$, computing its QR factorization $Y=QR$, and then setting $B = Q^\dagger A$. 

In the tensor-network setting, the matrix $A$ is never formed explicitly. Instead, it represents the TTNO--TTNS product. The random matrix $\Omega$ is also represented implicitly as a Khatri--Rao product of local random matrices,
\begin{equation}
    \Omega =\Omega^{(1)} \odot \Omega^{(2)} \odot \cdots \odot \Omega^{(L)} ,
\end{equation}
where $\odot$ denotes the columnwise Kronecker product. For each site $j$ in the tensor network, a random matrix $\Omega^{(j)}$ of size $d_j \times \chi$ is independently drawn, where $d_j$ is the local physical dimension and $\chi$ is the target bond dimension of the compressed output state. Here $\chi$ serves as the sketching dimension $p$ introduced for the matrix case in~\autoref{eq:matrix random compression}; in tensor networks, it simultaneously sets the bond dimension of the compressed TTNS, and we therefore denote it by $\chi$ throughout (including \autoref{tab:scalings} and Appendix~\ref{sec: scaling}). If
 $\Omega^{(j)} = \begin{bmatrix}
    c^{(j)}_1 & \cdots & c^{(j)}_\chi
    \end{bmatrix}$,
then the global sketching matrix has columns $\Omega = [(c_1^{(1)} \otimes \cdots \otimes c_1^{(n)}) \cdots (c_\chi^{(1)} \otimes \cdots \otimes c_\chi^{(L)})]$. This structure is illustrated in \autoref{fig:random tensor}.

\begin{figure}[htb]
{
\begin{tikzpicture}[
  state/.style={circle, draw, fill=purple!70, ultra thick, minimum size=6mm, inner sep=0pt},
  operator/.style={regular polygon,regular polygon sides=4, draw, fill=teal!30, ultra thick, minimum size=8mm, inner sep=0pt},
  gate/.style={draw, ellipse, fill=cyan!50!teal!90, ultra thick, minimum width=6mm, minimum height=14mm},
  gate2/.style={draw, ellipse, fill=cyan!40!black!50, ultra thick, minimum width=6mm, minimum height=14mm},
  >=Stealth,
]
\tikzmath{\dx = 1; \dy =-0.7; \lenarrow = 1.5; \d1=5.5*\dx; \d2=11.1*\dx; \dxx=12;\dyy=3.5*\dy;\dxt=5;} 
\node[state] (n1) at (0,0)  {}; \node[state] (n2) at (\dx,0)  {}; \node (ndot) at (2*\dx,0)  {$\cdots$}; \node[state] (n4) at (3*\dx,0)  {};
\node[circle, minimum size=1.5mm, ultra thick, draw, inner sep=0pt] (n5) at (1.5*\dx, \dy) {};
\draw (n1.south |- 0,-0.7)-- (n5.west); \draw (n5.east)-- (n4.south |- 0,-0.7);
\foreach \n in {n1,n2,n4}
  \draw (\n.south) -- (\n.south |- 0,-0.7);
\foreach \n in {n1,n2,n4}
  \draw (\n.north) -- (\n.north |- 0,0.7);
\draw (n5.south) -- ++(0,-0.4);
\end{tikzpicture}}
\caption{Khatri--Rao product of matrices. Each column of $\Omega$ is the Kronecker product of the corresponding columns of the local matrices $\Omega^{(j)}$.}
\label{fig:random tensor}
\end{figure}
\begin{figure}[htb]
{
\begin{tikzpicture}[
  state/.style={circle, draw, fill=cyan!70, ultra thick, minimum size=6mm, inner sep=0pt},
  state2/.style={circle, draw, fill=purple!70, ultra thick, minimum size=6mm, inner sep=0pt},
  state3/.style={regular polygon,regular polygon sides=3, draw, fill=red!30, ultra thick, minimum size=8mm, inner sep=0pt},
  state4/.style={circle, draw, fill=red!70, ultra thick, minimum size=6mm, inner sep=0pt},
  operator/.style={regular polygon,regular polygon sides=4, draw, fill=teal!30, ultra thick, minimum size=8mm, inner sep=0pt},
  gate/.style={draw, ellipse, fill=cyan!50!teal!90, ultra thick, minimum width=6mm, minimum height=14mm},
  gate2/.style={draw, ellipse, fill=cyan!40!black!50, ultra thick, minimum width=6mm, minimum height=14mm},
  gate3/.style={draw, ellipse, fill=cyan!50!teal!90, ultra thick, minimum width=14mm, minimum height=6mm},
  >=Stealth,
]
\tikzmath{\dx = 1; \dy =-1; \lenarrow = 0.6; \d1=3.5*\dx; \d2=7*\dx; \dxx=12;\dyy=5*\dy;\dxt=5; \dyyy=10*\dy;} 
\node (A) at (0, 0.7) {\textbf{(1)}};
\node[state] (A1t) at (0,0)  {}; \node[state] (A2t) at (\dx,0)  {}; \node[state] (A3t) at (2*\dx,0)  {};
\node[operator] (A1b) at (0,\dy) {}; \node[operator] (A2b) at (\dx,\dy) {}; \node[operator] (A3b) at (2*\dx,\dy) {};
\draw[very thick] (A1t)--(A2t)--(A3t) (A1b)--(A2b)--(A3b)
      (A1t)--(A1b) (A2t)--(A2b) (A3t)--(A3b);
\foreach \n in {A1b,A2b,A3b} {
  \draw[very thick] (\n.south) -- ++(0,-3mm);
}
\node (AmatA) at(1, -2.1) {$\mathbf{A}$};
\draw[->, very thick] (2.3*\dx,0.5*\dy) -- ++(\lenarrow,0); 

\node (B) at (\d1, 0.7) {\textbf{(2)}};
\draw[line width=0mm, gray!20, fill=gray!20] (\d1-0.4, -1.5) rectangle (\d1+1.5,-3);
\node[state] (B0t) at (\d1,0) {}; \node[state] (B1t) at (\d1+\dx,0)  {}; \node[state] (B2t) at (\d1+2*\dx,0)  {};
\node[operator] (B0b) at (\d1,\dy) {}; \node[operator] (B1b) at (\d1+\dx,\dy) {}; \node[operator] (B2b) at (\d1+2*\dx,\dy) {};
\node[state2] (B0r) at (\d1,2*\dy) {}; \node[state2] (B1r) at (\d1+\dx,2*\dy)  {}; 
\draw[very thick] (B0t)--(B1t) (B1t)--(B2t) (B0b)--(B1b) (B1b)--(B2b) (B0t)--(B0b) (B1t)--(B1b) (B2t)--(B2b) (B0r)--(B0b) (B1r)--(B1b);
\draw[very thick] (B2b.south) --++(0, -3mm);
\node[circle, minimum size=1.5mm, ultra thick, draw, inner sep=0pt] (n) at (\d1+0.5*\dx,2.5*\dy) {};
\draw ($(B0r.south)+(- 0,-0.2)$)-- (n.west); 
\draw (n.east)-- ($(B1r.south)+(- 0,-0.2)$);
\foreach \n in {B0r,B1r}
  \draw (\n.south) -- ++(0,-0.2);
\draw (n.south) -- ++(0,-0.4);
\node (Bphi) at (\d1+1.3, -2.7) {$\mathbf{\Omega}$};
\draw[->, very thick] (\d1+2.5*\dx,-0.5) -- node[midway, above] {} ++(\lenarrow,0); 
\node (C) at (\d2, 0.7) {\textbf{(3)}};
\node[gate2] (G0) at (\d2,-0.5) {};
\node[state] (C1t) at (\d2+1*\dx,0)  {};
\node[operator] (C1b) at (\d2+1*\dx,\dy) {};
\draw[very thick](C1t)--(C1b) ;
\draw[very thick] ($(G0.east)+(-0.1, 0.5)$) -- (C1t.west)
      ($(G0.east)+(-0.1,-0.5)$) -- (C1b.west);

\foreach \n in {C1b,G0} {
  \draw[very thick] (\n.south) -- ++(0,-3mm);
}

\draw[->, very thick] (\d2+1.5*\dx,-0.5) -- node[midway, above] {} ++(\lenarrow,0);

\node (D) at (\d2+2.6, 0.7) {\textbf{(4)}};
\node[gate3] (G1) at (\d2+3*\dx,0) {};
\draw[very thick] (G1.south west) --++(0, -0.4) (G1.south east) --++(0, -0.4);
\node (apro) at (\d2+3*\dx,-1) {$\approx$};
\node[state4] (D1n) at (\d2+2.5*\dx,-1.5) {};
\node[state3] (D2n) at (\d2+3.5*\dx,-1.5) {};
\draw[very thick] (D1n)--(D2n) (D1n.south)--++(0, -3mm) (D2n.south)--++(0, -3mm);

\draw[->, very thick] (\d2+3.7*\dx,-0.5) -- node[midway, above] {} ++(\lenarrow,0);
\draw[line width=0mm, gray!20, fill=gray!20] (\d2+4.5, 0.4) rectangle (\d2+7.5,-3);
\node (E) at (\d2+5, 0.7) {\textbf{(5)}}; 
\node (matB1) at (\d2+5,-2.7) {$\mathit{\mathbf{B^{(1)}}}$};
\node (matQ1) at (\d2+8, -2.7) {$\mathit{\mathbf{Q^{(1)}}}$};
\node[state] (E1t) at (\d2+5,0)  {}; \node[state] (E2t) at (\d2+5+\dx,0)  {}; \node[state] (E3t) at (\d2+5+2*\dx,0)  {};
\node[operator] (E1b) at (\d2+5,\dy) {}; \node[operator] (E2b) at (\d2+5+\dx,\dy) {}; \node[operator] (E3b) at (\d2+5+2*\dx,\dy) {};
\node[state3, rotate=180] (E1r) at (\d2+5+2*\dx, 2*\dy) {};
\node[state3, rotate=90] (E2r) at (\d2+5+3*\dx, 2*\dy) {};
\draw[very thick] (E1t)--(E2t)--(E3t) (E1b)--(E2b)--(E3b)
      (E1t)--(E1b) (E2t)--(E2b) (E3t)--(E3b) (E1r)--(E3b);
\draw[very thick] ($(E2r.south)+(0,-0.0)$) -- ($(E2r.south)+(0.3,-0.0)$)-- ($(E2r.south)+(0.3,-0.6)$);
\foreach \n in {E1b,E2b,E3b} {
  \draw[very thick] (\n.south) -- ++(0,-3mm);
}
\draw[very thick] (E1r.north) to[out=-90,in=-135,looseness=1] (\d2+5+2.3*\dx, 2.5*\dy) to[out=45,in=180,looseness=1] (\d2+5+2.4*\dx, 2*\dy) to[out=100,in=0] (E2r.north);

\node (F) at (0, 0.7+\dyy) {\textbf{(6)}};
\draw[line width=0mm, gray!20, fill=gray!20] (-0.3, 0.4+\dyy) rectangle (2.5,-3+\dyy);
\node (FmatB1) at (0.3, -2.7+\dyy) {$\mathbf{B^{(1)}}$};
\node[state] (F1t) at (0,\dyy)  {}; \node[state] (F2t) at (\dx,\dyy)  {}; \node[gate] (F3t) at (2*\dx,\dyy-0.5)  {};
\node[operator] (F1b) at (0,\dy+\dyy) {}; \node[operator] (F2b) at (\dx,\dy+\dyy) {};
\draw[very thick] (F1t)--(F2t)--($(F3t.west)+(0.1,0.5)$) (F1b)--(F2b)--($(F3t.west)+(0.1,-0.5)$)
      (F1t)--(F1b) (F2t)--(F2b) ;
\foreach \n in {F1b,F2b,F3t} {
  \draw[very thick] (\n.south) -- ++(0,-3mm);
}

\draw[->, very thick] (2.5*\dx,0.5*\dy+\dyy) -- ++(\lenarrow,0); 

\node (G) at (\d1, 0.7+\dyy) {\textbf{(7)}};
\node[state] (G1t) at (\d1,\dyy)  {}; \node[state] (G2t) at (\dx+\d1,\dyy)  {}; \node[gate] (G3t) at (2*\dx+\d1,\dyy-0.5)  {};
\node[operator] (G1b) at (0+\d1,\dy+\dyy) {}; \node[operator] (G2b) at (\dx+\d1,\dy+\dyy) {}; 

\node[state2] (G1r) at (\d1,2*\dy+\dyy) {}; 
\node[circle, minimum size=1.5mm, ultra thick, draw, inner sep=0pt] (Gn) at (\d1,2.5*\dy+\dyy) {};
\draw[very thick] (G1t)--(G2t)--($(G3t.west)+(0.1,0.5)$) (G1b)--(G2b)--($(G3t.west)+(0.1,-0.5)$) (G1t)--(G1b)--(G1r)--(Gn) (G2t)--(G2b);
\foreach \n in {G1b,G2b,G3t,Gn} {
  \draw[very thick] (\n.south) -- ++(0,-3mm);
}
\draw[->, very thick] (2.5*\dx+\d1,0.5*\dy+\dyy) -- ++(\lenarrow,0); 

\node (H) at (\d2, 0.7+\dyy) {\textbf{(8)}};
\node[gate2] (H0) at (\d2,-0.5+\dyy) {};
\node[state] (H1t) at (\d2+1*\dx,+\dyy)  {};
\node[operator] (H1b) at (\d2+1*\dx,\dy+\dyy) {};
\node[gate2] (H2) at (\d2+2*\dx,-0.5+\dyy) {};
\draw[very thick](H1t)--(H1b) ;
\draw[very thick] ($(H0.east)+(-0.1, 0.5)$) -- (H1t) -- ($(H2.west)+(0.1, 0.5)$)
      ($(H0.east)+(-0.1,-0.5)$) -- (H1b) -- ($(H2.west)+(0.1,-0.5)$);

\foreach \n in {H1b,H0,H2} {
  \draw[very thick] (\n.south) -- ++(0,-3mm);
}
\draw[->, very thick] (\d2+2.5*\dx,0.5*\dy+\dyy) -- ++(\lenarrow,0); 

\node (I) at (\d2+3.7, 0.7+\dyy) {\textbf{(9)}};
\node[gate3] (I1) at (\d2+4*\dx,\dyy) {};
\draw[very thick] (I1.south west) --++(0, -0.4) (I1.south) --++(0, -0.4) (I1.south east) --++(0, -0.4);
\node (apro) at (\d2+4*\dx,-1+\dyy) {$\approx$};
\node[state4] (I1n) at (\d2+3.5*\dx,-1.5+\dyy) {};
\node[state3] (I2n) at (\d2+4.5*\dx,-1.5+\dyy) {};
\draw[very thick] (I1n)--(I2n) (I1n.south)--++(0, -3mm)  ($(I2n.south) + (-0.1, 0)$)--++(0, -3mm) ($(I2n.south) + (0.1, 0)$)--++(0, -3mm);

\draw[->, very thick] (\d2+4.8*\dx,-0.5+\dyy) -- node[midway, above] {} ++(\lenarrow,0);
\draw[line width=0mm, gray!20, fill=gray!20] (\d2+5.5, 0.4+\dyy) rectangle (\d2+8.5,-1.5+\dyy);
\draw[line width=0mm, gray!20, fill=gray!20] (\d2+5.5, -1.5+\dyy) rectangle (\d2+8,-3+\dyy);
\node (J) at (\d2+6, 0.7+\dyy) {\textbf{(10)}}; 
\node (JmatB2) at (\d2+6, -2.7+\dyy){$\mathit{\mathbf{B^{(2)}}}$};
\node (JmatQ2) at (\d2+8.5, -2.7+\dyy){$\mathit{\mathbf{Q^{(2)}}}$};
\node (JmatQ1) at (\d2+9.5, -2.7+\dyy){$\mathit{\mathbf{Q^{(1)}}}$};
\node[state] (J1t) at (\d2+6,\dyy)  {}; \node[state] (J2t) at (\d2+6+\dx,\dyy)  {}; \node[gate2] (J3t) at (\d2+6+2*\dx,\dyy-0.5)  {};
\node[operator] (J1b) at (\d2+6,\dy+\dyy) {}; \node[operator] (J2b) at (\d2+6+\dx,\dy+\dyy) {};
\node[state3, rotate=180] (J1r) at (\d2+5.5+2*\dx, 2*\dy+\dyy) {};
\node[state3, rotate=90] (J2r) at (\d2+5.5+3*\dx, 2*\dy+\dyy) {};
\node[state3, rotate=90] (J3r) at (\d2+5.5+4*\dx, 2*\dy+\dyy) {};
\draw[very thick] (J1t)--(J2t)--($(J3t.west)+(0.1, 0.5)$) (J1b)--(J2b)--($(J3t.west)+(0.1, -0.5)$) (J1t)--(J1b) (J2t)--(J2b);   
\draw[very thick] ($(J2r.south)+(0,-0.2)$) -- ($(J2r.south)+(0.3,-0.2)$)-- ($(J2r.south)+(0.3,-0.6)$);
\draw[very thick] ($(J3r.south)+(0,-0.0)$) -- ($(J3r.south)+(0.3,-0.0)$)-- ($(J3r.south)+(0.3,-0.6)$);
\foreach \n in {J1b} {
  \draw[very thick] (\n.south) -- ++(0,-3mm);
}

\draw[very thick] (J1r.north) to[out=-90,in=-135,looseness=1] (\d2+5.5+2.3*\dx, 2.5*\dy+\dyy) to[out=45,in=180,looseness=1] (\d2+5.5+2.4*\dx, 2*\dy+\dyy) to[out=100,in=0] (J2r.north);
\draw[very thick] (J2b.south) to [out=-90,in=90,looseness=1] (\d2+6+\dx,\dy+\dyy-0.5) to [out=-90, in=120] ($(J1r.south)+(-0.2,0)$);
\draw[very thick] (J3t.south) to [out=-90,in=90,looseness=1] (\d2+6+2*\dx,\dy+\dyy-0.5) to [out=-90, in=60] ($(J1r.south)+(0.2,0)$);
\draw[very thick] ($(J2r.south)+(0,0.2)$) to [out=0,in=180,looseness=1] (\d2+5.5+3.5*\dx, 2*\dy+\dyy) to [out=0, in=270] (J3r.north);

\node (K) at (0, 0.7+\dyyy) {\textbf{(11)}};
\node[state,fill=red!20] (K1) at (0, \dyyy-0.5) {};
\node[state3,rotate=90] (K2)at (\dx, \dyyy-0.5) {};
\node[state3,rotate=90] (K3) at (2*\dx, \dyyy-0.5) {};
\draw[very thick] (K1)-- (K2.north);
\draw[very thick] ($(K2.south)+(0,0.2)$) to [out=0,in=180,looseness=1] (\dx+0.5, \dyyy-0.5) to [out=0, in=270] (K3.north);
\draw[very thick] ($(K2.south)+(0,-0.2)$) -- ($(K2.south)+(0.3,-0.2)$)-- ($(K2.south)+(0.3,-0.6)$);
\draw[very thick] ($(K3.south)+(0,-0.0)$) -- ($(K3.south)+(0.3,-0.0)$)-- ($(K3.south)+(0.3,-0.6)$);
\draw[very thick] (K1.south) --++(0,-0.3);

\node (approx) at (\d1-0.5, \dyyy-0.5) {$\mathbf{\approx}$};

\node[state] (L0t) at (\d1+0.3,\dyyy) {}; \node[state] (L1t) at (\d1+0.3+\dx,\dyyy)  {}; \node[state] (L2t) at (\d1+0.3+2*\dx,\dyyy)  {};
\node[operator] (L0b) at (\d1+0.3,\dy+\dyyy) {}; \node[operator] (L1b) at (\d1+0.3+\dx,\dy+\dyyy) {}; \node[operator] (L2b) at (\d1+0.3+2*\dx,\dy+\dyyy) {};

\draw[very thick] (L0t)--(L1t) (L1t)--(L2t) (L0b)--(L1b) (L1b)--(L2b) (L0t)--(L0b) (L1t)--(L1b) (L2t)--(L2b) ;
\foreach \n in {L0b,L1b, L2b}
  \draw (\n.south) -- ++(0,-0.2);
\end{tikzpicture}
}
\caption{Illustration of the successive randomized compression (SRC) method. Circle tensors represent the state, square tensors represent the operator, ellipses are intermediate tensors, and triangle tensors are orthogonal tensors.}
\label{fig:src}
\end{figure}

Operationally, one traverses the tree from the leaves toward the root. At each node $i$, the neighboring subtrees that have already been processed provide environment blocks. These blocks are obtained by contracting the corresponding TTNO and TTNS tensors with the local random matrices $\Omega^{(j)}$, as illustrated in~\autoref{fig:src} steps (2) and (3). The resulting local sketch plays the role of $Y=A\Omega$ in the matrix formulation. It is then orthogonalized to produce the compressed tensor $Q^{(i)}$, which becomes the new tensor at node $i$; see \autoref{fig:src} step(4). In this way, the compressed basis is constructed without explicitly forming the enlarged TTNO--TTNS tensor at that node.

Next, by contracting the conjugate transpose $Q^{(i)\dagger}$ with the TTNO and TTNS tensors, we obtain the matrix $B^{(i)}$. The same sketching-and-compression procedure is then applied recursively to this residual network. Once the traversal reaches the root, all remaining tensors are contracted directly. The resulting tensor network provides an approximation to the product of the TTNO and TTNS.

\subsubsection{Variational Fitting}
\label{subsubsec:variational_fitting}

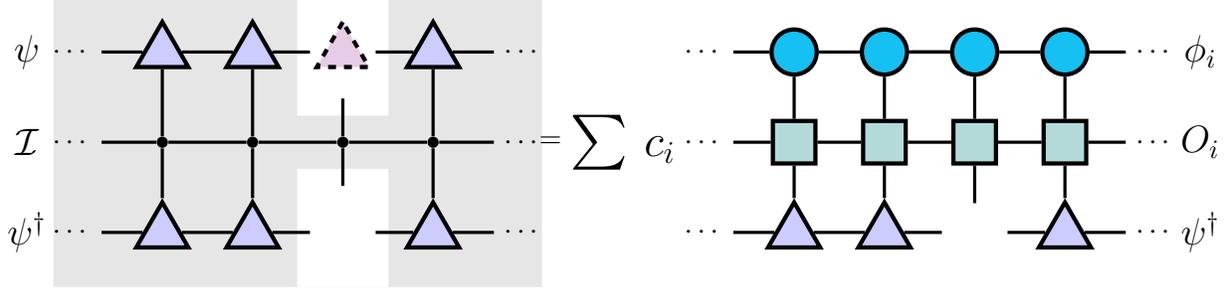
\begin{figure}[htb]
\begin{tikzpicture}[
  x=12mm, y=12mm, >=Stealth,
  given/.style={circle, draw, ultra thick, fill= cyan!70, minimum size=6mm, inner sep=0pt},
  operator/.style={regular polygon,regular polygon sides=4, draw, ultra thick, fill=teal!30, minimum size=8mm, inner sep=0pt},
  unknown/.style={regular polygon,regular polygon sides=3, draw, ultra thick, fill=blue!20, minimum size=8mm, inner sep=0pt},
]

\draw[line width=0mm, gray!20, fill=gray!20] (-0.2,0.6) rectangle (5.2,-2.6);
\draw[line width=0mm, white, fill=white] (2.5,0.6) rectangle (3.5,-0.7);
\draw[line width=0mm, white, fill=white] (2.5,-1.3) rectangle (3.5,-2.6);
\node at (-0.5,0) {\large$\psi$};
\node (cdot1) at (0,0)   {$\cdots$};
\node[unknown] (k1t) at (1,0) {};
\node[unknown] (k2t) at (2,0) {};
\node[unknown] (k3t) at (4,0) {};
\node[unknown,dashed,fill=violet!20] (knt) at (3,0) {};
\node (cdot2) at (5,0)   {$\cdots$};

\node at (-0.5,-1) {\large$\mathcal{I}$};

\node (cdot3) at (0,-1)   {$\cdots$};
\node (dot1) at (1,-1) [circle,fill,inner sep=1.5pt]{};
\node (dot2) at (2,-1) [circle,fill,inner sep=1.5pt]{};
\node (dot3) at (3,-1) [circle,fill,inner sep=1.5pt]{};
\node (dot4) at (4,-1) [circle,fill,inner sep=1.5pt]{};
\node (cdot4) at (5,-1)   {$\cdots$};

\node at (-0.5,-2) {\large$\psi^\dagger$};
\node (cdot5) at (0,-2)   {$\cdots$};
\node[unknown] (k1b) at (1,-2) {};
\node[unknown] (k2b) at (2,-2) {};
\node[unknown] (k3b) at (4,-2) {};
\node (cdot6) at (5,-2)   {$\cdots$};

\draw[very thick] (cdot1)--(k1t)--(k2t) (k3t)--(cdot2)
(cdot3)--(dot1)--(dot2)--(dot3)--(dot4)--(cdot4) (cdot5)--(k1b)--(k2b) (k3b)--(cdot6) (k1t)--(dot1)--(k1b) (k2t)--(dot2)--(k2b) (k3t)--(dot4)--(k3b);
\draw[very thick] (k2t.east) -- ++(5mm,0);
\draw[very thick] (k3t.west) -- ++(-5mm,0);
\draw[very thick] (dot3.north) -- ++(0,5mm);
\draw[very thick] (dot3.south) -- ++(0,-5mm);
\draw[very thick] (k2b.east) -- ++(5mm,0);
\draw[very thick] (k3b.west) -- ++(-5mm,0);

\node (equal) at (5.3,-1) {$=$};
\node (sum) at (6.1, -1) {\LARGE $\sum$ \Large $c_i$ };
\node (cdot7) at (7,0)   {$\cdots$};
\node[given] (r1t) at (8,0) {};
\node[given] (r2t) at (9,0) {};
\node[given] (r3t) at (10,0) {};
\node[given] (r4t) at (11,0) {};
\node (cdot8) at (12,0)   {$\cdots$};
\node (ket) at (12.5,0) {\large$\phi_i$};

\node (cdot9) at (7,-1)   {$\cdots$};
\node [operator](sq1) at (8,-1) {};
\node [operator](sq2) at (9,-1) {};
\node [operator](sq3) at (10,-1) {};
\node [operator](sq4) at (11,-1) {};
\node (cdot10) at (12,-1)   {$\cdots$};
\node (op) at (12.5,-1) {\large$O_i$};

\node (cdot11) at (7,-2)   {$\cdots$};
\node[unknown] (r1b) at (8,-2) {};
\node[unknown] (r2b) at (9,-2) {};
\node[unknown] (r3b) at (11,-2) {};
\node (cdot12) at (12,-2)   {$\cdots$};
\node (bra) at (12.5,-2) {\large$\psi^\dagger$};

\draw[very thick] (cdot7)--(r1t)--(r2t)--(r3t)--(r4t)--(cdot8)
(cdot9)--(sq1)--(sq2)--(sq3)--(sq4)--(cdot10) 
(cdot11)--(r1b)--(r2b) (r3b)--(cdot12) (r1t)--(sq1)--(r1b) (r2t)--(sq2)--(r2b) (r3t)--(sq3) (r4t)--(sq4)--(r3b);
\draw[very thick] (r2t.east) -- ++(5mm,0);
\draw[very thick] (r3t.west) -- ++(-5mm,0);
\draw[very thick] (sq3.south) -- ++(0,-5mm);
\draw[very thick] (r2b.east) -- ++(5mm,0);
\draw[very thick] (r3b.west) -- ++(-5mm,0);
\end{tikzpicture}
\caption{Variational fitting. Circle tensors are the given state $\phi_i$, square tensors are the given operator $O_i$, and triangle tensors are the target state $\psi$. The tensors in the shaded area on the left represent the effective Hamiltonian $A$. The dashed triangle tensor is the solution of the linear system $A x = b$. The right-hand side represents the vector $b$ after contraction.}
\label{fig:variational}
\end{figure}

Variational fitting is a method that can be used to compress the sum of multiple TTNSs, or multiple TTNOs applied to TTNSs, into a single TTNS $\psi$~\cite{Rakhuba2016, baiardi2019optimization, Rano2025}, i.e.,
\begin{align}
    \psi \approx \sum_i c_i O_i {\phi_i},
\end{align}
where the $O_i$ are TTNOs, ${\phi_i}$ are TTNSs, and $c_i$ are complex coefficients. When the $O_i$ are identity operators, the aforementioned equation is equivalent to compressing the sum of multiple weighted TTNSs. The idea behind this compression is to find the best fit by minimizing the difference between the compressed and uncompressed TTNSs:
\begin{equation}
    \min_{\psi} \norm*{ \psi - \sum_i c_i O_i {\phi_i}}_2
\end{equation}
by iteratively updating the $\psi$ tensors. At each node $j$, we construct an effective Hamiltonian $H_{\text{eff},j}$ by contracting our current estimate of the TTNS $\psi$ with its complex conjugate $\psi^*$ except for the tensor $\psi_j$ at node $j$ and its complex conjugate counterpart. Then we need to solve the linear system
\begin{equation}
    H_{\text{eff,}j} y_j = \sum_i c_i O_{\text{eff},i,j} \phi _{i,j},
\end{equation}
where $y_j$ is the solution tensor at the current site, $O_{i,j}$ is the local TTNO tensor at the current site, and $\phi _{i,j}$ is the local TTNS tensor at the current site. A graphical version of this equation for MPS is depicted in \autoref{fig:variational}. The above linear system of equations is solved using the MINRES algorithm from SciPy~\cite{2020SciPy-NMeth}. The obtained solution tensor $y_j$ then updates the current $\psi_j$ tensor. Since we perform one-site updates at a time, the virtual bond dimension of the solution tensor network state will not change. To enable dynamic bond dimension updates, we further compute the local residuals after solving the linear system
\begin{equation}
    r = H_{\text{eff}, j} y_j - \sum_i c_i O_{i,j} \phi_{i,j}.
\end{equation}
If the norm of this residual vector is larger than a given threshold, we apply a rank-revealing QR decomposition and expand the solution $y_j$ with the obtained $Q$.

The ability to compress sums of TTNSs makes variational fitting useful inside the LOBPCG method. We will next consider the alternating least squares approach and its adaptation to TTNs within the LOBPCG eigenvalue solver.

\subsubsection{Alternating least squares (ALS) method}
\label{subsubsec:als}

\begin{figure}[htb]
{
\begin{tikzpicture}
    [
  x=12mm, y=12mm, >=Stealth,
  given/.style={circle, draw, ultra thick, fill= cyan!70, minimum size=6mm, inner sep=0pt},
  operator/.style={regular polygon,regular polygon sides=4, draw, ultra thick, fill=teal!30, minimum size=8mm, inner sep=0pt},
  unknown/.style={regular polygon,regular polygon sides=3, draw, ultra thick, fill=blue!20, minimum size=8mm, inner sep=0pt},
]

\draw[line width=0mm, gray!20, fill=gray!20] (-0.2,0.6) rectangle (5.2,-2.6);
\draw[line width=0mm, white, fill=white] (2.5,0.6) rectangle (3.5,-0.6);
\draw[line width=0mm, white, fill=white] (2.5,-1.4) rectangle (3.5,-2.6);

\node at (-0.5,0) {\large$\psi$};
\node (cdot1) at (0,0)   {$\cdots$};
\node[unknown] (k1t) at (1,0) {};
\node[unknown] (k2t) at (2,0) {};
\node[unknown] (k3t) at (4,0) {};
\node[unknown,dashed,fill=violet!20] (knt) at (3,0) {};
\node (cdot2) at (5,0)   {$\cdots$};

\node at (-0.5,-1) {\large${H}$};
\node (cdot3) at (0,-1)   {$\cdots$};
\node[operator] (dot1) at (1,-1) {};
\node[operator] (dot2) at (2,-1) {};
\node[operator] (dot3) at (3,-1) {};
\node[operator] (dot4) at (4,-1) {};
\node (cdot4) at (5,-1)   {$\cdots$};

\node at (-0.5,-2) {\large$\psi^\dagger$};
\node (cdot5) at (0,-2)   {$\cdots$};
\node[unknown] (k1b) at (1,-2) {};
\node[unknown] (k2b) at (2,-2) {};
\node[unknown] (k3b) at (4,-2) {};
\node (cdot6) at (5,-2)   {$\cdots$};

\draw[very thick] (cdot1)--(k1t)--(k2t) (k3t)--(cdot2)
(cdot3)--(dot1)--(dot2)--(dot3)--(dot4)--(cdot4) (cdot5)--(k1b)--(k2b) (k3b)--(cdot6) (k1t)--(dot1)--(k1b) (k2t)--(dot2)--(k2b) (k3t)--(dot4)--(k3b);
\draw[very thick] (k2t.east) -- ++(5mm,0);
\draw[very thick] (k3t.west) -- ++(-5mm,0);
\draw[very thick] (dot3.north) -- ++(0,5mm);
\draw[very thick] (dot3.south) -- ++(0,-5mm);
\draw[very thick] (k2b.east) -- ++(5mm,0);
\draw[very thick] (k3b.west) -- ++(-5mm,0);

\node (Ax) at (2.5, -3) {$Ax$};
\node (b) at (9.5, -3) {$b$};
\node (equal) at (6,-1) {$=$};

\node (cdot7) at (7,0)   {$\cdots$};
\node[given] (r1t) at (8,0) {};
\node[given] (r2t) at (9,0) {};
\node[given] (r3t) at (10,0) {};
\node[given] (r4t) at (11,0) {};
\node (cdot8) at (12,0)   {$\cdots$};
\node (bra) at (12.5,0) {\large$\phi$};

\node (cdot9) at (7,-1)   {$\cdots$};
\node (sq1) at (8,-1) [circle,fill,inner sep=1.5pt]{};
\node (sq2) at (9,-1) [circle,fill,inner sep=1.5pt]{};
\node (sq3) at (10,-1) [circle,fill,inner sep=1.5pt]{};
\node (sq4) at (11,-1) [circle,fill,inner sep=1.5pt]{};
\node (cdot10) at (12,-1)   {$\cdots$};
\node (bra) at (12.5,-1) {\large$\mathcal{I}$};

\node (cdot11) at (7,-2)   {$\cdots$};
\node[unknown] (r1b) at (8,-2) {};
\node[unknown] (r2b) at (9,-2) {};
\node[unknown] (r3b) at (11,-2) {};
\node (cdot12) at (12,-2)   {$\cdots$};
\node (bra) at (12.5,-2) {\large$\psi^\dagger$};

\draw[very thick] (cdot7)--(r1t)--(r2t)--(r3t)--(r4t)--(cdot8)
(cdot9)--(sq1)--(sq2)--(sq3)--(sq4)--(cdot10) (cdot11)--(r1b)--(r2b) (r3b)--(cdot12) (r1t)--(sq1)--(r1b) (r2t)--(sq2)--(r2b) (r3t)--(sq3) (r4t)--(sq4)--(r3b);
\draw[very thick] (r2t.east) -- ++(5mm,0);
\draw[very thick] (r3t.west) -- ++(-5mm,0);
\draw[very thick] (sq3.south) -- ++(0,-5mm);
\draw[very thick] (r2b.east) -- ++(5mm,0);
\draw[very thick] (r3b.west) -- ++(-5mm,0);
\end{tikzpicture}
}
\caption{Alternating least squares. Circle tensors are the given state $\phi$, square tensors are the given operator $H$, and triangle tensors are the target state $\psi$. The tensors in the shaded area on the left represent the effective Hamiltonian $A$. The dashed triangle tensor is the solution of the linear system $A x = b$. The right-hand side represents the vector $b$ after contraction.}
\label{fig:als}
\end{figure}
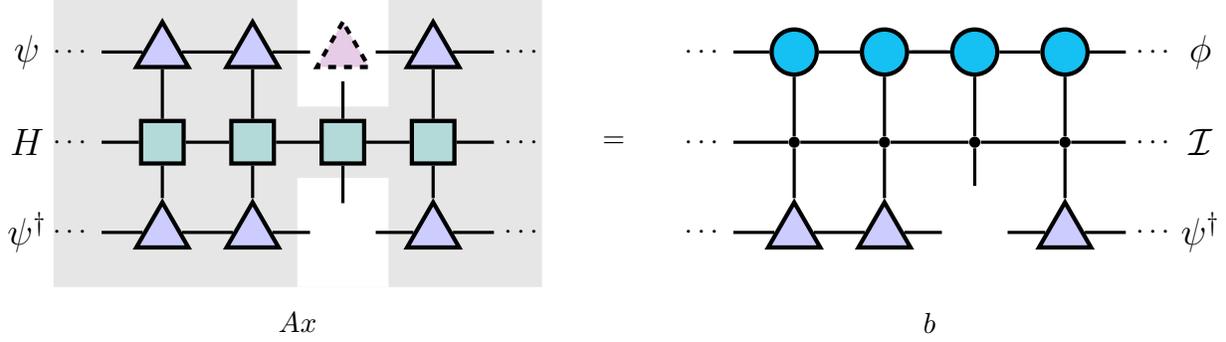

To solve the linear system $H \psi = \phi$ in a tree tensor network framework, i.e., where $H$ is represented as a TTNO and $\psi$ and $\phi$ are TTNSs, we implement the alternating least squares (ALS) method. ALS is an iterative method to solve local linear system equations. Similar to variational fitting, we update our solution node by node. For each node, we have a local, smaller system of linear equations to solve. When updating node $\phi_{j}$, we construct the local effective Hamiltonian $H_{\text{eff},j}$ by contracting all other nodes, effectively computing $\bra{\psi}H\ket{\psi}$, except for the current node $\phi_j$ and its conjugate counterpart. The effective right-hand side vector resembles the contraction of $\langle\psi |\phi\rangle$ but leaving out $\psi_j^*$. A graphical depiction of the equation to solve is shown in \autoref{fig:als}. Minres can find the local best fit for the current node, denoted $\Tilde{\psi}_j$. We then replace the old node tensor $\psi_j$ with the new solution tensor $\Tilde{\psi}_j$ and move to the next node. This iteration over all nodes in the tree constitutes one sweep; multiple sweeps are typically required for convergence. Since this method is only used for approximating the preconditioned residual vector in LOBPCG, we do not need to fully converge the solution. In practice, running two sweeps is typically sufficient to obtain an accurate approximation. For the inverse iteration method, we can increase the accuracy by performing more sweeps. The difference between variational fitting and ALS is in how the local effective Hamiltonian is constructed. In variational fitting, the local effective Hamiltonian is constructed by contracting the bra $\psi$ and the ket $\psi$ tensors with an identity operator, i.e., $\langle\psi|I|\psi\rangle$, whereas in ALS, the local effective Hamiltonian is constructed by contracting $\langle \psi|H|\psi\rangle$.

\section{Results}
\label{sec:results}

In the following, we assess the numerical performance and accuracy of our TTN framework for two benchmark systems --- 64-dimensional coupled oscillators and acetonitrile~\cite{Carbonniere2004, Leclerc2014} --- and compare runtime and accuracy across different tree structures (MPS, T-tree, leaf-only tree~\cite{Larsson2019}, fork-$n$ tree, and the full T3NS~\cite{gunst2018t3ns}). The corresponding tree structures are described in Appendix~\ref{sec:appendix_tree_structure}. All computational codes and results are publicly available on GitHub (\url{https://github.com/sunshuo987/vibcal}). Benchmark calculations were carried out on an Apple M4 processor with 10 cores and 16 GB of RAM; however, the actual simulations required substantially less memory.

\subsection{64-dimensional coupled oscillators}
\label{subsec:64D_coupled_oscillators}

The vibrational Hamiltonian of bilinearly coupled oscillators is described by
\begin{equation}
    \mathrm{H} = \sum_{i=1}^{D} \frac{\omega_i}{2}\left(-\frac{\partial^2}{\partial q_i^2}+ q_i^2\right) + \sum_{j=1}^{D-1} \sum_{i>j}^D \alpha_{ij} q_i q_j.\label{eq:coupled_harmonic_oscillator}
\end{equation}
We set $D = 64$, $\omega _j = \sqrt{\frac{j}{2}}$, and $\alpha_{ij} = 0.1$, same as Refs.~\onlinecite{Thomas2015, Rakhuba2016, Zhang2024}. These parameters create a moderately coupled system that tests the ability of TTN methods to handle bilinear interactions. The resulting Hamiltonian has $2080$ unique terms ($64$ single-mode terms together with $\binom{64}{2}=2016$ bilinear couplings), which nonetheless yields an optimized TTNO with a maximum bond dimension of only $3$ for both the MPS and T-tree topologies. In this work, we use Hermite basis functions for which the kinetic term in~\autoref{eq:coupled_harmonic_oscillator} is given by~\cite{baye1986generalised, dolgov2014computation}
\begin{align}
T_{mn} = 
\begin{cases}
(-1)^{\,m-n}\!\left(\dfrac{2}{(x_m-x_n)^2}-\dfrac{1}{2}\right), & m \neq n, \\
\dfrac{1}{6} \left(4N - 1 - 2 x_m^2\right), & m = n,
\end{cases}
\end{align}
and $q_m$ is a diagonal matrix with diagonal entries $(x_1, x_2, \dots, x_k)$, where $x_i$ are the roots of the Hermite polynomials.

\begin{figure}[htb]
  \centering
  \subfloat[\label{fig:64d_convergence}]{%
    \includegraphics[width=0.5\linewidth]{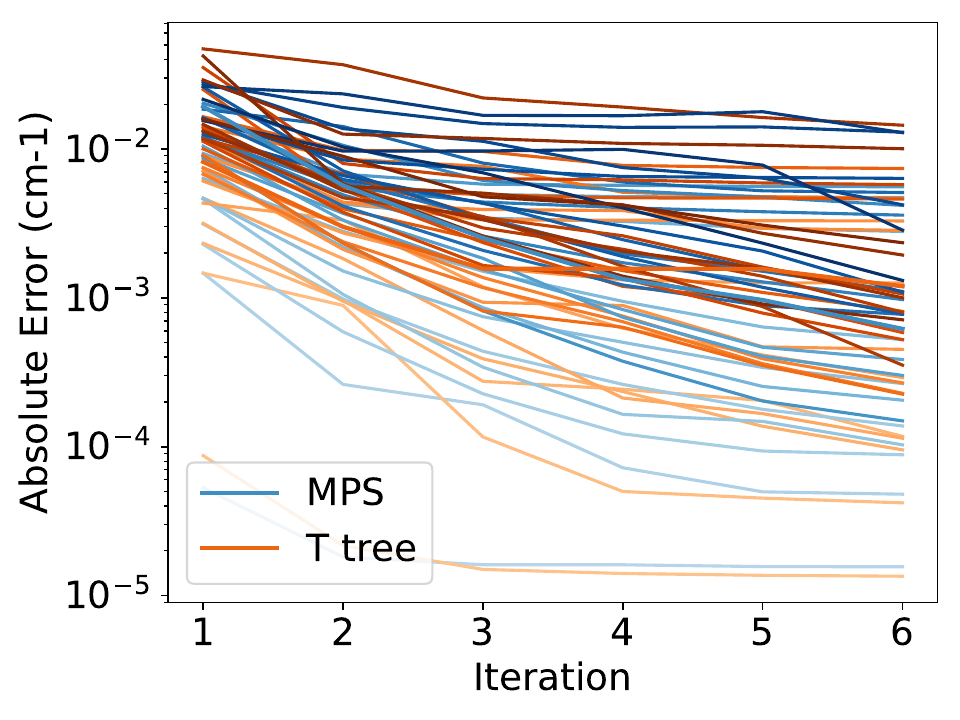}}\hfill
  \subfloat[\label{fig:64d_errors}]{%
    \includegraphics[width=0.5\linewidth]{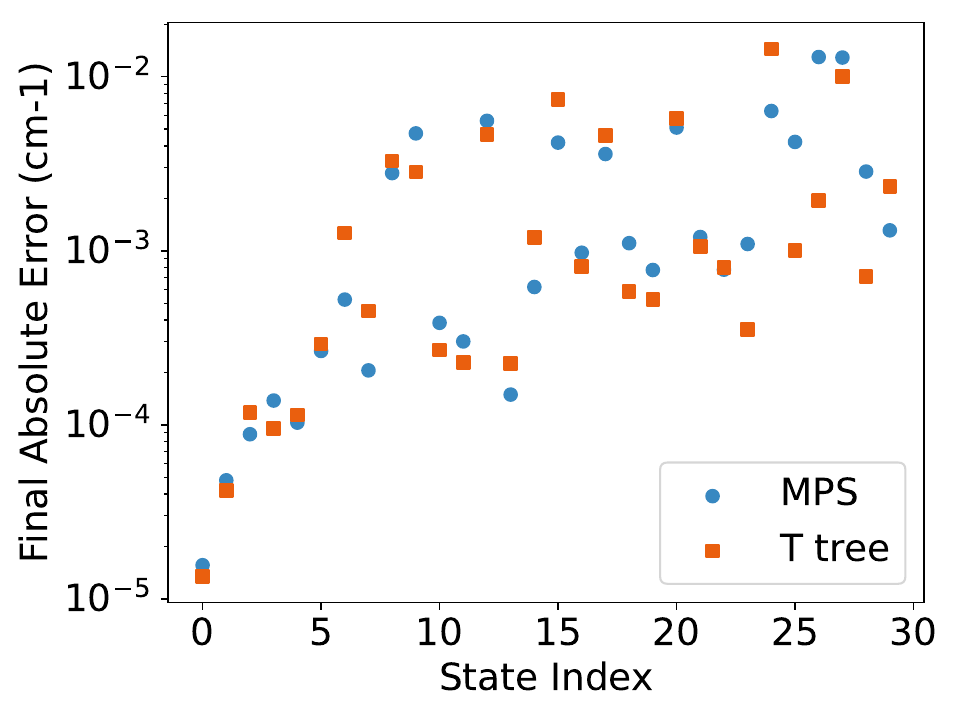}}
  \caption{(a) Convergence curves over six iterations for the lowest 30 eigenstates of the 64-dimensional coupled-oscillator model, computed using the MPS and T tree ansatz. Color intensity encodes the energy level, with lighter shades corresponding to lower energies.
(b) Final absolute error for both structures.}
  \label{fig:both}
\end{figure}

We represent both the state and the Hamiltonian using MPS and T tree structures, as illustrated in Appendix \ref{sec:appendix_tree_structure} \autoref{fig:ttn_structures_64}. Each oscillator is assigned a physical basis size of 8, and the maximum virtual bond dimension of the states is set to 5. We compute the lowest 30 eigenstates using the LOBPCG algorithm. Because exact energy levels are available analytically for this model, it provides a clean benchmark of the quality of the tensor-network ansatz at fixed bond dimension. 
The corresponding absolute errors are plotted in \autoref{fig:both}. Both MPS and T tree achieve errors below $10^{-2}$~cm$^{-1}$ for low-lying states; however, for many states T tree yields smaller errors than MPS. This result demonstrates the improved accuracy of tree tensor network representations and highlights their potential advantages over one-dimensional MPS for capturing correlations in vibrational Hamiltonians.

\subsection{Acetonitrile ($\mathrm{CH_3CN}$)}
\label{subsubsec:acetonitrile}

\begin{figure}[htbp]
  \centering
  \subfloat[\label{fig:ch3cn_modes}]{%
    \includegraphics[width=0.7\linewidth]{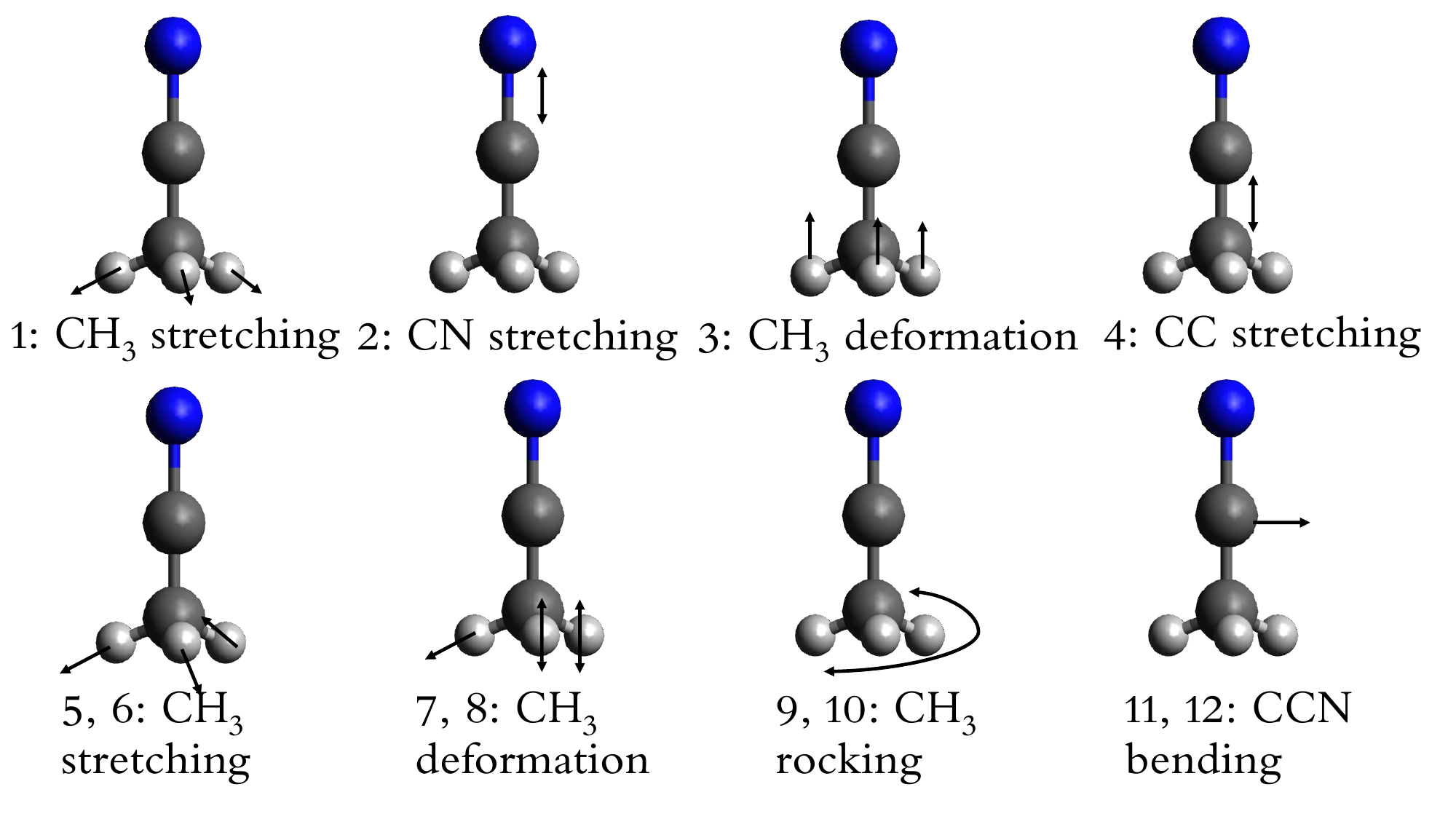}}\hfill
  \subfloat[\label{fig:ch3cn_errors}]{%
    \includegraphics[width=0.9\linewidth]{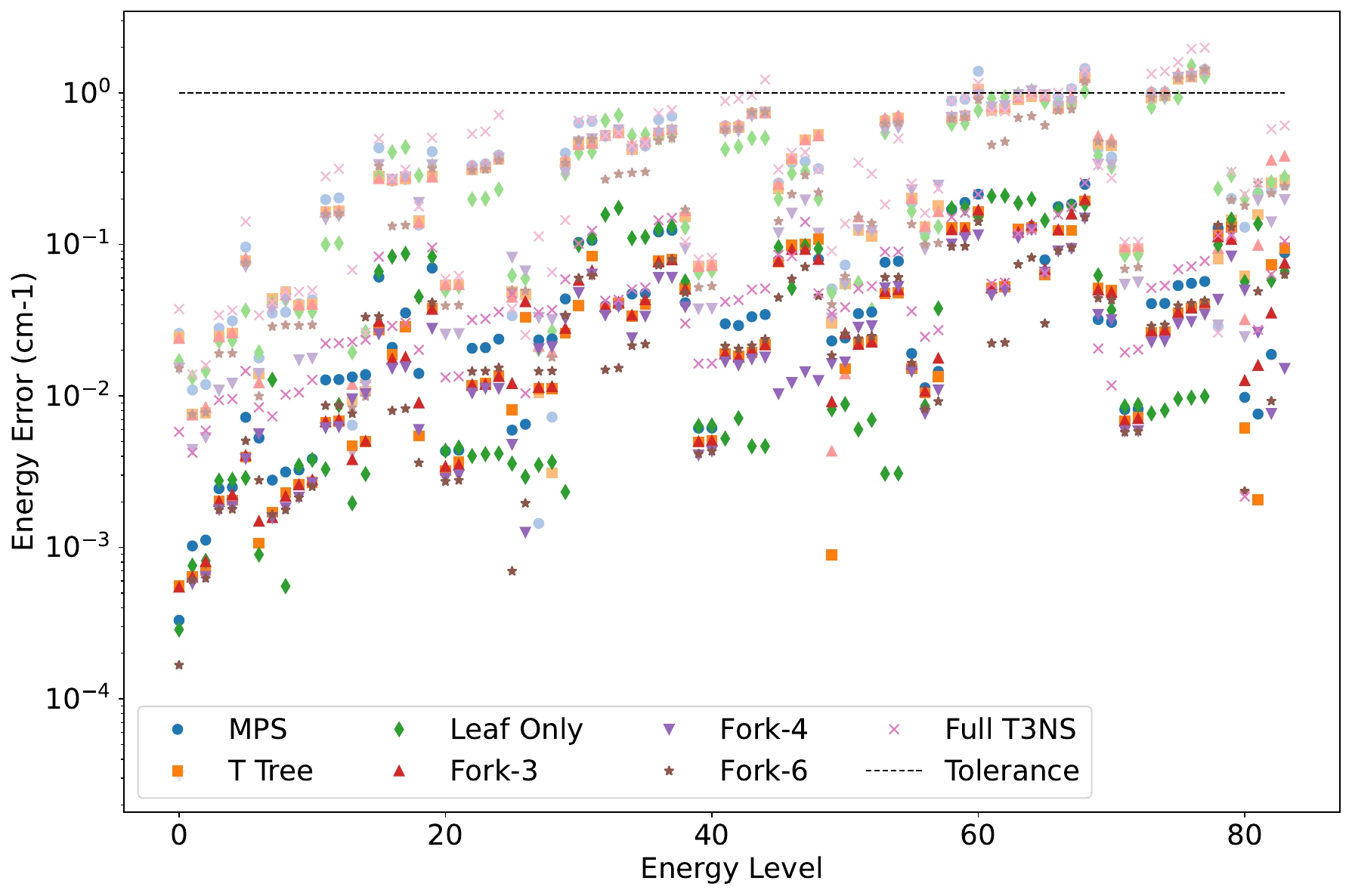}}
	\caption{(a) Motion of the 12 vibrational modes of acetonitrile ($\mathrm{CH_3CN}$). (b) Final absolute error of the 84 eigenstates computed via MPS, T-tree, leaf-only, fork-3, fork-4, fork-6, and full T3NS tensor network structures. Lighter-colored markers correspond to the LOBPCG results, whereas darker markers show the inverse-iteration results obtained using the LOBPCG output as the initial state.}
	\label{fig:ch3cn}
\end{figure}

Acetonitrile is a molecule with 12 vibrational modes (see \autoref{fig:ch3cn_modes}) that has been studied extensively in recent years~\cite{Carbonniere2004, Avila2011, Leclerc2014, Rakhuba2016, Larsson2019, larsson2025benchmarking}. The $C_{3v}$ symmetry of acetonitrile results in degenerate vibrational modes and strong anharmonic coupling between them, which poses significant challenges for standard eigensolvers. In this work, we employed the conventional local basis sizes, which are $\{ 9, 7, 9, 9, 9, 9, 7, 7, 9, 9, 27, 27 \}$, and adopted the same coupling strengths reported in the literature~\cite{Begue2005,Avila2011, Zhang2024}. The vibrational Hamiltonian of acetonitrile used in these references has the form
\begin{align}
    \mathrm{H} = -\frac{1}{2} \sum_i \omega_i \frac{\partial^2}{\partial q_i^2} + \frac{1}{2} \sum_{i} \omega_i q_i^2 + \frac{1}{6} \sum_{i,j,k} k_{i,j,k}^{(3)} q_i q_j q_k + \frac{1}{24} \sum_{i,j,k,\ell} k_{i,j,k,\ell}^{(4)} q_i q_j q_k q_\ell,
\end{align}
where $k^{(3)}$ and $k^{(4)}$ are the cubic and quartic force constants, and indices $i$, $j$, $k$, $\ell$ represent the 12 vibrational modes. This Hamiltonian has $311$ unique terms ($12$ harmonic, $108$ cubic, and $191$ quartic terms). Representing it as an optimized TTNO yields a maximum bond dimension of only $18$--$25$ depending on the tree topology (e.g.\ $18$ for the T-tree and fork-3 structures and $25$ for the full T3NS).

We first applied the LOBPCG method to compute the lowest 84 eigenstates with a maximum virtual bond dimension of 12. The preconditioner shift in \autoref{eq:precondition} was set to $s=9$ for all states. To obtain a sufficiently accurate preconditioned residual, we solved \autoref{eq:solve precondition} by ALS using two sweeps, and the maximum number of Minres iterations per local solve was set to 10. For all states, we ran five outer LOBPCG iterations. These states were then used as initial guesses for inverse iteration in order to obtain higher-accuracy solutions. For the refinement stage, the maximum virtual bond dimension was increased to 20, and we again performed five outer iterations. States whose Rayleigh quotients differed by less than 5~cm$^{-1}$ were grouped into the same block. We solved \autoref{eq:inverse iteration update} with ALS using one sweep per outer iteration, and the maximum number of Minres iterations per local solve was set to 10\,000 to ensure the local optimization converges (often convergence is reached with far fewer iterations). After each iteration, the shift $s$ in \autoref{eq:inverse iteration update} was updated to the mean Rayleigh quotient of the states in the corresponding block. These settings were chosen to expose topology-dependent trends at moderate computational cost rather than to exhaust the variational space fully; larger bond dimensions or tighter inner solves would reduce the errors further. Summary statistics are collected in \autoref{tab:ch3cn_summary}, while level-resolved signed errors are listed in \autoref{tab:ch3cn_lobpcg_energy} and \autoref{tab:ch3cn_ii_energy}. As reference, we use the MPS energies with a maximum virtual bond dimension of 100 from Ref.~\onlinecite{Larsson2019}. Although more accurate values are available in Ref.~\onlinecite{larsson2025benchmarking}, the larger physical basis used there makes a direct comparison less consistent with our setup. 

Comparisons between MPS and leaf-only tree topologies have been reported previously, for instance for acetonitrile in Ref.~\onlinecite{Larsson2019} and in a different context in Ref.~\onlinecite{Li2025}. In the present work, we extend these comparisons to a broader range of structures within a single, consistent code base. Specifically, we systematically compare seven tensor network topologies: MPS, T-tree, leaf-only tree~\cite{Larsson2019}, fork-3 tree~\cite{Bauernfeind2017, Chepiga2019}, fork-4 tree, fork-6 tree, and the full T3NS tree~\cite{gunst2018t3ns}. Fork-$n$ trees~ \cite{Bauernfeind2017} (also known as comb tensor networks~\cite{Chepiga2019}) consist of $n$ auxiliary backbone nodes connected in a chain, each of which supports a chain of physical nodes hanging off it. The corresponding structures are shown in Appendix~\ref{sec:appendix_tree_structure}, \autoref{fig:ttn_structures_ch3cn_1}, and \autoref{fig:ttn_structures_ch3cn_2}. Note that the physical basis sizes employed here are not sufficient to reach the sub-0.1~cm$^{-1}$ accuracy reported in Ref.~\onlinecite{larsson2025benchmarking}. With larger basis size, more highly branched leaf-only tree topologies can become more favorable in terms of runtime as discussed in Appendix~\ref{sec: scaling}. 

The lighter symbols in \autoref{fig:ch3cn_errors} show that LOBPCG alone already yields a qualitatively correct spectrum across all seven topologies. The errors grow with energy level, as expected from the increasing density of states and stronger mixing among excited configurations. Several higher states remain near or above the 1~cm$^{-1}$ tolerance line, but the mean absolute errors are already in the range 0.298--0.436~cm$^{-1}$. Among the LOBPCG calculations, fork-6 gives the smallest mean absolute error (0.298~cm$^{-1}$), followed closely by fork-4 (0.330~cm$^{-1}$) and the leaf-only tree (0.332~cm$^{-1}$). The fastest topologies are MPS and T-tree. Relative to fork-4, the leaf-only tree is about 30~\% slower and fork-6 about 44~\% slower, which makes the accuracy--cost trade-off already visible at the LOBPCG stage. This advantage of branched trees is especially visible in congested manifolds. For example, in the 1756--1760~cm$^{-1}$ window (levels 32--37), the fork-6 errors are about 0.27--0.50~cm$^{-1}$, compared with about 0.44--0.70~cm$^{-1}$ for MPS. A similar pattern appears in the 2111--2123~cm$^{-1}$ region (levels 58--68), where fork-6 consistently lowers the LOBPCG errors relative to the linear and weakly branched topologies.

The darker symbols in \autoref{fig:ch3cn_errors} demonstrate the effect of inverse-iteration refinement. For every topology and every one of the 84 states, the final error falls below 1~cm$^{-1}$. The worst remaining errors are only about 0.15--0.26~cm$^{-1}$, depending on the topology, and the mean errors decrease by roughly one order of magnitude for all trees. Fork-4 yields the smallest mean absolute error (0.026~cm$^{-1}$) and, together with fork-6, the largest number of best-matched states in the state-by-state comparison. Because fork-6 requires about 58~\% more runtime than fork-4, and the leaf-only tree needs roughly twice as much time as fork-4, the fork-4 topology offers the best overall balance between accuracy and computational cost in the present benchmark. Moreover, the errors within nearly degenerate level pairs are usually very similar, which indicates that the block solvers treat the corresponding invariant subspaces consistently. The high-lying 2119--2123~cm$^{-1}$ manifold remains one of the more challenging parts of the spectrum even after refinement, yet fork-4 and fork-6 keep the errors there near 0.02--0.15~cm$^{-1}$, whereas MPS remains in the broader 0.05--0.25~cm$^{-1}$ range. Thus, the topology dependence is not washed out by inverse iteration; rather, refinement preserves the advantage of the more suitable trees while reducing the errors for all topologies.

In both \autoref{tab:ch3cn_lobpcg_energy} and \autoref{tab:ch3cn_ii_energy}, some energies appear negative. This does not indicate higher accuracy than the reference values, but rather arises because certain states are not strictly orthogonal, with overlaps below 0.02. This artifact can be removed by solving the generalized eigenvalue problem $Hc = ESc$, where $S$ is the overlap matrix, as in Ref.~\onlinecite{larsson2025benchmarking}.  T-tree and fork-3 display very similar accuracies, which is natural because contracting the two side auxiliary nodes of fork-3 into its central node yields a T-tree. The T-tree nevertheless remains faster because it contains fewer tensors to optimize. The leaf-only tree used here corresponds to the optimized topology reported in Ref.~\onlinecite{Larsson2019}.
However, this optimization is performed for a single target state and therefore does not necessarily generalize to other energy levels. This likely explains why the optimized leaf-only tree achieves adequate accuracy overall, while not showing a clear advantage over the other topologies and tending to require longer runtimes due to its larger number of nodes. The unoptimized full T3NS is less competitive than the other branched topologies in this particular setup, but it still reaches a mean error of 0.054~cm$^{-1}$ after inverse iteration, confirming the robustness of the eigensolver framework across rather different trees. The level-resolved data also show that no single topology is uniformly optimal: different states prefer different trees, which is consistent with the fact that different vibrational excitations activate different correlation patterns. For instance, the leaf-only tree performs well for overtones of $\nu_{11}$, likely because it was optimized for the 10th excitation level ($3\nu_{11}$). Conversely, fork-4 and fork-6 are more effective for the $\nu_{5,7,9} + n\nu_{11}$ levels, as these structures provide closer connectivity between those specific modes. Even so, the overall convergence across all seven topologies is encouraging.

\begin{table}[ht]
\caption{Summary of accuracy and runtime for the acetonitrile calculations. The column ``Number of best-matched states'' counts how many states attain the smallest absolute error for a given topology; ties are counted for all matching topologies.}
\label{tab:ch3cn_summary}

\begin{tabular}{>{\centering\arraybackslash}p{2cm}| >{\centering\arraybackslash}p{2cm}  >{\centering\arraybackslash}p{5cm}  >{\centering\arraybackslash}p{4cm}  >{\centering\arraybackslash}p{2.5cm}}
\toprule
Method & Tree Type& {Number of best-matched states} & Mean Error (cm$^{-1}$)& Wall-clock Time (s)\\
\hline
\multirow[t]{7}{*}{LOBPCG} & MPS & 2 & 0.379 & 647.176 \\
 & T-Tree & 2 & 0.346 & 704.564 \\
 & Leaf-only & 21 & 0.332 & 1602.970 \\
 & Fork-3 & 6 & 0.355 & 957.105 \\
 & Fork-4 & 23 & 0.330 & 1231.541 \\
 & Fork-6 & 22 & 0.298 & 1775.155 \\
 & Full T3NS & 7 & 0.436 & 1268.846 \\
\cline{1-5}
\multirow[t]{7}{*}{\raisebox{-0.5\height}{\shortstack[c]{\strut Inverse \\ [-5pt] Iteration}}} & MPS & 5 & 0.048 & 1946.084 \\
 & T-Tree & 12 & 0.037 & 2478.980 \\
 & Leaf-only & 30 & 0.052 & 7432.548 \\
 & Fork-3 & 8 & 0.037 & 2900.775 \\
 & Fork-4 & 34 & 0.026 & 3527.168 \\
 & Fork-6 & 34 & 0.030 & 5591.586 \\
 & Full T3NS & 4 & 0.054 & 5232.781 \\
\cline{1-5}
\hline
\end{tabular}
\end{table}

\begin{longtable}{p{2.3cm}|p{2cm}|p{1.5cm}p{1.5cm}p{1.8cm}p{1.5cm}p{1.5cm}p{1.5cm}p{1.9cm}}
\caption{Reference zero-point vibrational energy (level 0) and excitation energies relative to the zero-point vibrational energy (levels 1--83), together with signed errors (in cm$^{-1}$) for acetonitrile obtained with LOBPCG at virtual bond dimension 12.}
\label{tab:ch3cn_lobpcg_energy}\\

\toprule
Level & Ref~\cite{Larsson2019}(cm$^{-1}$) & MPS & T Tree & Leaf-Only & Fork-3 & Fork-4 & Fork-6 & Full T3NS \\
\hline
\endfirsthead

\multicolumn{9}{c}{\tablename\ \thetable\ -- continued from previous page}\\
\toprule
Level & Ref~\cite{Larsson2019}(cm$^{-1}$) & MPS & T Tree & Leaf Only & Fork-3 & Fork-4 & Fork-6 & Full T3NS \\
\hline
\endhead
\multicolumn{9}{r}{continued on next page}\\
\endfoot
\endlastfoot
0 (ZPE) & 9837.407 & 0.026 & 0.024 & 0.017 & 0.024 & \best{0.015} & \best{0.015} & 0.037 \\
1 ($\nu_{11}$) & 360.990 & 0.011 & 0.008 & 0.013 & 0.007 & \best{0.004} & 0.008 & 0.014 \\
2 ($\nu_{11}$) & 360.990 & 0.012 & 0.008 & 0.015 & 0.008 & \best{0.005} & 0.008 & 0.016 \\
3 ($2\nu_{11}$) & 723.179 & 0.028 & 0.025 & 0.023 & 0.025 & \best{0.011} & 0.019 & 0.034 \\
4 ($2\nu_{11}$) & 723.179 & 0.031 & 0.026 & 0.023 & 0.026 & \best{0.012} & 0.019 & 0.037 \\
5 ($2\nu_{11}$) & 723.825 & 0.096 & 0.078 & \best{0.037} & 0.081 & 0.072 & 0.075 & 0.142 \\
6 ($\nu_{4}$) & 900.659 & 0.018 & 0.014 & 0.019 & 0.012 & 0.014 & \best{0.010} & 0.034 \\
7 ($\nu_{9}$) & 1034.124 & 0.035 & 0.044 & 0.041 & 0.041 & 0.037 & \best{0.029} & 0.041 \\
8 ($\nu_{9}$) & 1034.124 & 0.036 & 0.049 & 0.043 & 0.047 & 0.044 & \best{0.029} & 0.049 \\
9 ($3\nu_{11}$) & 1086.552 & 0.041 & 0.040 & 0.036 & 0.040 & \best{0.017} & 0.029 & 0.049 \\
10 ($3\nu_{11}$) & 1086.552 & 0.043 & 0.041 & 0.036 & 0.040 & \best{0.018} & 0.029 & 0.050 \\
11 ($3\nu_{11}$) & 1087.774 & 0.198 & 0.165 & \best{0.100} & 0.164 & 0.147 & 0.158 & 0.282 \\
12 ($3\nu_{11}$) & 1087.774 & 0.203 & 0.167 & \best{0.102} & 0.167 & 0.154 & 0.160 & 0.316 \\
13 ($\nu_{4} + \nu_{11}$) & 1259.809 & 0.006 & 0.009 & 0.019 & 0.012 & \best{0.004} & 0.008 & -0.068 \\
14 ($\nu_{4} + \nu_{11}$) & 1259.809 & 0.014 & 0.011 & 0.026 & 0.013 & 0.012 & \best{0.010} & 0.026 \\
15 ($\nu_{3}$) & 1388.967 & 0.434 & 0.281 & 0.284 & \best{0.272} & 0.337 & 0.329 & 0.499 \\
16 ($\nu_{9} + \nu_{11}$) & 1394.679 & 0.262 & 0.266 & 0.407 & 0.270 & 0.268 & \best{0.132} & 0.274 \\
17 ($\nu_{9} + \nu_{11}$) & 1394.679 & 0.269 & 0.271 & 0.439 & 0.281 & 0.278 & \best{0.134} & 0.312 \\
18 ($\nu_{9} + \nu_{11}$) & 1394.898 & 0.134 & 0.143 & 0.285 & 0.139 & 0.189 & \best{0.044} & 0.178 \\
19 ($\nu_{9} + \nu_{11}$) & 1397.680 & 0.410 & 0.282 & 0.331 & \best{0.279} & 0.336 & 0.318 & 0.506 \\
20 ($4\nu_{11}$) & 1451.093 & 0.055 & 0.054 & 0.050 & 0.054 & \best{0.026} & 0.039 & 0.059 \\
21 ($4\nu_{11}$) & 1451.093 & 0.055 & 0.054 & 0.053 & 0.054 & \best{0.026} & 0.040 & 0.062 \\
22 ($4\nu_{11}$) & 1452.818 & 0.332 & 0.311 & \best{0.199} & 0.330 & 0.313 & 0.310 & 0.537 \\
23 ($4\nu_{11}$) & 1452.818 & 0.341 & 0.320 & \best{0.201} & 0.337 & 0.321 & 0.313 & 0.557 \\
24 ($4\nu_{11}$) & 1453.394 & 0.389 & 0.367 & \best{0.231} & 0.381 & 0.370 & 0.359 & 0.716 \\
25 ($\nu_{7}$) & 1483.219 & 0.034 & -0.049 & -0.062 & -0.045 & -0.082 & -0.050 & \best{-0.038} \\
26 ($\nu_{7}$) & 1483.220 & 0.047 & -0.047 & -0.060 & -0.042 & -0.067 & -0.049 & \best{0.025} \\
27 ($\nu_{4} + 2\nu_{11}$) & 1620.199 & 0.001 & \best{-0.011} & 0.020 & -0.021 & -0.033 & -0.023 & -0.113 \\
28 ($\nu_{4} + 2\nu_{11}$) & 1620.199 & 0.007 & \best{-0.003} & 0.027 & -0.019 & -0.032 & -0.018 & -0.066 \\
29 ($\nu_{4} + 2\nu_{11}$) & 1620.744 & 0.401 & 0.349 & 0.294 & 0.360 & 0.298 & 0.344 & \best{0.145} \\
30 ($\nu_{3} + \nu_{11}$) & 1749.519 & 0.633 & 0.466 & \best{0.402} & 0.457 & 0.476 & 0.489 & 0.656 \\
31 ($\nu_{3} + \nu_{11}$) & 1749.519 & 0.650 & 0.472 & \best{0.409} & 0.466 & 0.498 & 0.496 & 0.666 \\
32 ($\nu_{9} + 2\nu_{11}$) & 1756.412 & 0.521 & 0.526 & 0.659 & 0.525 & 0.519 & \best{0.269} & 0.524 \\
33 ($\nu_{9} + 2\nu_{11}$) & 1756.412 & 0.544 & 0.546 & 0.713 & 0.552 & 0.571 & \best{0.292} & 0.547 \\
34 ($\nu_{9} + 2\nu_{11}$) & 1757.119 & 0.442 & 0.426 & 0.523 & 0.454 & 0.422 & \best{0.297} & 0.450 \\
35 ($\nu_{9} + 2\nu_{11}$) & 1757.119 & 0.447 & 0.470 & 0.533 & 0.483 & 0.447 & \best{0.301} & 0.469 \\
36 ($\nu_{9} + 2\nu_{11}$) & 1759.760 & 0.667 & 0.527 & 0.503 & 0.562 & 0.544 & \best{0.485} & 0.736 \\
37 ($\nu_{9} + 2\nu_{11}$) & 1759.760 & 0.704 & 0.570 & 0.527 & 0.578 & 0.563 & \best{0.502} & 0.770 \\
38 ($2\nu_{4}$) & 1785.177 & -0.149 & -0.153 & -0.130 & -0.168 & \best{-0.094} & -0.170 & -0.104 \\
39 ($5\nu_{11}$) & 1816.786 & 0.069 & 0.072 & 0.065 & 0.072 & \best{0.038} & 0.052 & 0.080 \\
40 ($5\nu_{11}$) & 1816.786 & 0.070 & 0.073 & 0.067 & 0.073 & \best{0.038} & 0.053 & 0.081 \\
41 ($5\nu_{11}$) & 1818.938 & 0.609 & 0.585 & \best{0.425} & 0.604 & 0.555 & 0.572 & 0.885 \\
42 ($5\nu_{11}$) & 1818.939 & 0.611 & 0.594 & \best{0.443} & 0.615 & 0.561 & 0.583 & 0.916 \\
43 ($5\nu_{11}$) & 1820.016 & 0.739 & 0.733 & \best{0.503} & 0.733 & 0.704 & 0.749 & 0.970 \\
44 ($5\nu_{11}$) & 1820.016 & 0.751 & 0.748 & \best{0.504} & 0.741 & 0.740 & 0.751 & 1.227 \\
45 ($\nu_{7} + \nu_{11}$) & 1844.245 & 0.254 & 0.235 & 0.200 & 0.250 & \best{0.120} & 0.143 & 0.313 \\
46 ($\nu_{7} + \nu_{11}$) & 1844.316 & 0.346 & 0.366 & 0.295 & 0.373 & \best{0.161} & 0.214 & 0.402 \\
47 ($\nu_{7} + \nu_{11}$) & 1844.317 & 0.353 & 0.491 & 0.305 & 0.491 & \best{0.197} & 0.287 & 0.407 \\
48 ($\nu_{7} + \nu_{11}$) & 1844.676 & 0.317 & 0.530 & 0.200 & 0.523 & \best{0.119} & 0.220 & 0.311 \\
49 ($\nu_{4} + \nu_{9}$) & 1931.514 & 0.051 & 0.030 & 0.048 & \best{0.004} & 0.033 & 0.040 & 0.091 \\
50 ($\nu_{4} + \nu_{9}$) & 1931.514 & 0.073 & 0.055 & 0.056 & \best{0.014} & 0.055 & 0.062 & 0.137 \\
51 ($\nu_{4} + 3\nu_{11}$) & 1981.815 & -0.123 & -0.124 & \best{-0.056} & -0.152 & -0.126 & -0.150 & -0.346 \\
52 ($\nu_{4} + 3\nu_{11}$) & 1981.815 & -0.114 & -0.114 & \best{-0.037} & -0.131 & -0.123 & -0.138 & -0.293 \\
53 ($\nu_{4} + 3\nu_{11}$) & 1982.818 & 0.667 & 0.651 & 0.549 & 0.678 & 0.567 & 0.627 & \best{0.183} \\
54 ($\nu_{4} + 3\nu_{11}$) & 1982.818 & 0.693 & 0.681 & 0.615 & 0.708 & 0.592 & 0.638 & \best{0.501} \\
55 ($2\nu_{9}$) & 2057.044 & 0.188 & 0.202 & 0.166 & 0.201 & 0.230 & \best{0.136} & 0.252 \\
56 ($2\nu_{9}$) & 2065.265 & 0.121 & 0.131 & 0.112 & 0.132 & \best{0.093} & 0.100 & 0.161 \\
57 ($2\nu_{9}$) & 2065.265 & 0.171 & 0.181 & 0.131 & 0.165 & 0.246 & \best{0.102} & 0.234 \\
58 ($\nu_{3} + 2\nu_{11}$) & 2111.364 & 0.887 & 0.676 & \best{0.624} & 0.703 & 0.691 & 0.678 & 0.886 \\
59 ($\nu_{3} + 2\nu_{11}$) & 2111.364 & 0.908 & 0.710 & \best{0.632} & 0.715 & 0.701 & 0.691 & 0.933 \\
60 ($\nu_{3} + 2\nu_{11}$) & 2112.281 & 1.391 & 1.059 & \best{0.768} & 1.005 & 0.973 & 0.902 & 1.162 \\
61 ($\nu_{9} + 3\nu_{11}$) & 2119.307 & 0.809 & 0.770 & 0.923 & 0.815 & 0.809 & \best{0.454} & 0.752 \\
62 ($\nu_{9} + 3\nu_{11}$) & 2119.307 & 0.815 & 0.788 & 0.941 & 0.818 & 0.826 & \best{0.477} & 0.757 \\
63 ($\nu_{9} + 3\nu_{11}$) & 2120.521 & 0.940 & 0.909 & 0.982 & 0.938 & 0.974 & \best{0.685} & 0.945 \\
64 ($\nu_{9} + 3\nu_{11}$) & 2120.521 & 0.968 & 0.951 & 1.032 & 1.017 & 1.036 & \best{0.703} & 0.966 \\
65 ($\nu_{9} + 3\nu_{11}$) & 2120.889 & 0.938 & 0.980 & 0.876 & 0.958 & 0.970 & \best{0.611} & 0.973 \\
66 ($\nu_{9} + 3\nu_{11}$) & 2122.816 & 0.946 & 0.794 & 0.828 & 0.806 & 0.818 & \best{0.766} & 1.002 \\
67 ($\nu_{9} + 3\nu_{11}$) & 2122.816 & 1.067 & 0.880 & 0.843 & 0.944 & 0.878 & \best{0.780} & 1.057 \\
68 ($\nu_{9} + 3\nu_{11}$) & 2123.282 & 1.451 & 1.272 & \best{1.023} & 1.265 & 1.322 & 1.191 & 1.424 \\
69 ($2\nu_{4} + \nu_{11}$) & 2142.444 & -0.430 & -0.459 & -0.387 & -0.520 & -0.338 & -0.478 & \best{-0.333} \\
70 ($2\nu_{4} + \nu_{11}$) & 2142.444 & -0.379 & -0.452 & -0.326 & -0.493 & -0.335 & -0.465 & \best{-0.274} \\
71 ($6\nu_{11}$) & 2183.617 & 0.089 & 0.093 & 0.085 & 0.096 & \best{0.055} & 0.069 & 0.104 \\
72 ($6\nu_{11}$) & 2183.617 & 0.092 & 0.093 & 0.085 & 0.098 & \best{0.056} & 0.071 & 0.104 \\
73 ($6\nu_{11}$) & 2186.117 & 1.012 & 0.926 & \best{0.809} & 1.018 & 0.928 & 0.934 & 1.338 \\
74 ($6\nu_{11}$) & 2186.117 & 1.018 & 0.956 & 0.943 & 1.026 & \best{0.933} & 0.952 & 1.389 \\
75 ($6\nu_{11}$) & 2187.618 & 1.282 & 1.235 & \best{0.932} & 1.345 & 1.266 & 1.249 & 1.596 \\
76 ($6\nu_{11}$) & 2187.618 & 1.308 & \best{1.271} & 1.510 & 1.376 & 1.295 & 1.275 & 1.953 \\
77 ($6\nu_{11}$) & 2188.119 & 1.432 & 1.375 & \best{1.279} & 1.437 & 1.372 & 1.431 & 1.990 \\
78 ($\nu_{7} + 2\nu_{11}$) & 2206.608 & 0.029 & 0.080 & 0.233 & 0.095 & -0.029 & 0.119 & \best{-0.026} \\
79 ($\nu_{7} + 2\nu_{11}$) & 2206.615 & 0.201 & 0.145 & 0.287 & \best{0.114} & 0.117 & 0.197 & 0.302 \\
80 ($\nu_{7} + 2\nu_{11}$) & 2206.757 & 0.130 & 0.062 & 0.200 & 0.032 & \best{0.025} & 0.180 & 0.215 \\
81 ($\nu_{7} + 2\nu_{11}$) & 2206.758 & 0.219 & 0.158 & 0.216 & \best{0.099} & 0.196 & 0.254 & 0.256 \\
82 ($\nu_{7} + 2\nu_{11}$) & 2207.541 & 0.225 & 0.255 & 0.257 & 0.360 & \best{0.141} & 0.218 & 0.577 \\
83 ($\nu_{7} + 2\nu_{11}$) & 2207.541 & 0.248 & 0.266 & 0.277 & 0.383 & \best{0.198} & 0.242 & 0.610
 \\
\end{longtable}

\begin{longtable}{p{2.3cm}|p{2cm}|p{1.5cm}p{1.5cm}p{1.8cm}p{1.5cm}p{1.5cm}p{1.5cm}p{1.9cm}}
\caption{Reference zero-point vibrational energy (level 0) and excitation energies relative to the zero-point vibrational energy (levels 1--83), together with signed errors (in cm$^{-1}$) for acetonitrile obtained with inverse iteration at virtual bond dimension 20.}
\label{tab:ch3cn_ii_energy}\\

\toprule
Level & Ref~\cite{Larsson2019}(cm$^{-1}$) & MPS & T Tree & Leaf-Only & Fork-3 & Fork-4 & Fork-6 & Full T3NS \\
\hline
\endfirsthead

\multicolumn{9}{c}{\tablename\ \thetable\ -- continued from previous page}\\
\toprule
Level & Ref~\cite{Larsson2019}(cm$^{-1}$) & MPS & T Tree & Leaf Only & Fork-3 & Fork-4 & Fork-6 & Full T3NS \\
\hline
\endhead
\multicolumn{9}{r}{continued on next page}\\
\endfoot
\endlastfoot
0 (ZPE) & 9837.407 & \best{0.000} & 0.001 & \best{0.000} & 0.001 & \best{0.000} & \best{0.000} & 0.006 \\
1 ($\nu_{11}$) & 360.990 & \best{0.001} & \best{0.001} & \best{0.001} & \best{0.001} & \best{0.001} & \best{0.001} & 0.004 \\
2 ($\nu_{11}$) & 360.990 & \best{0.001} & \best{0.001} & \best{0.001} & \best{0.001} & \best{0.001} & \best{0.001} & 0.006 \\
3 ($2\nu_{11}$) & 723.179 & \best{0.002} & \best{0.002} & 0.003 & \best{0.002} & \best{0.002} & \best{0.002} & 0.009 \\
4 ($2\nu_{11}$) & 723.179 & \best{0.002} & \best{0.002} & 0.003 & \best{0.002} & \best{0.002} & \best{0.002} & 0.010 \\
5 ($2\nu_{11}$) & 723.825 & 0.007 & 0.004 & \best{0.003} & 0.004 & 0.004 & 0.005 & 0.015 \\
6 ($\nu_{4}$) & 900.659 & 0.005 & \best{0.001} & \best{-0.001} & \best{0.001} & 0.006 & 0.003 & 0.008 \\
7 ($\nu_{9}$) & 1034.124 & 0.003 & \best{0.002} & -0.013 & \best{-0.002} & \best{0.002} & \best{0.002} & 0.007 \\
8 ($\nu_{9}$) & 1034.124 & 0.003 & 0.002 & \best{0.001} & 0.002 & 0.002 & 0.002 & 0.010 \\
9 ($3\nu_{11}$) & 1086.552 & 0.003 & 0.003 & 0.003 & 0.003 & \best{0.002} & \best{0.002} & 0.011 \\
10 ($3\nu_{11}$) & 1086.552 & 0.004 & \best{0.003} & 0.004 & \best{0.003} & \best{0.003} & \best{0.003} & 0.013 \\
11 ($3\nu_{11}$) & 1087.774 & 0.013 & 0.007 & \best{0.003} & 0.007 & 0.006 & 0.009 & 0.022 \\
12 ($3\nu_{11}$) & 1087.774 & 0.013 & 0.007 & 0.009 & 0.007 & \best{0.006} & 0.009 & 0.022 \\
13 ($\nu_{4} + \nu_{11}$) & 1259.809 & 0.013 & 0.005 & \best{0.002} & 0.004 & 0.010 & 0.008 & 0.023 \\
14 ($\nu_{4} + \nu_{11}$) & 1259.809 & 0.014 & 0.005 & \best{0.003} & 0.005 & 0.010 & 0.033 & 0.024 \\
15 ($\nu_{3}$) & 1388.967 & 0.061 & 0.027 & 0.066 & 0.031 & \best{0.026} & 0.033 & 0.083 \\
16 ($\nu_{9} + \nu_{11}$) & 1394.679 & 0.021 & 0.019 & 0.083 & 0.018 & 0.015 & \best{0.008} & 0.029 \\
17 ($\nu_{9} + \nu_{11}$) & 1394.679 & 0.035 & 0.029 & 0.087 & 0.018 & 0.016 & \best{0.008} & 0.030 \\
18 ($\nu_{9} + \nu_{11}$) & 1394.898 & -0.014 & -0.005 & 0.045 & 0.009 & 0.006 & \best{0.004} & 0.020 \\
19 ($\nu_{9} + \nu_{11}$) & 1397.680 & 0.070 & 0.038 & 0.083 & 0.037 & \best{0.028} & 0.041 & 0.095 \\
20 ($4\nu_{11}$) & 1451.093 & 0.004 & \best{0.003} & 0.004 & \best{0.003} & \best{0.003} & \best{0.003} & 0.013 \\
21 ($4\nu_{11}$) & 1451.093 & 0.004 & 0.004 & 0.005 & 0.004 & \best{0.003} & \best{0.003} & 0.013 \\
22 ($4\nu_{11}$) & 1452.818 & 0.021 & 0.012 & \best{0.004} & 0.012 & 0.010 & 0.014 & 0.032 \\
23 ($4\nu_{11}$) & 1452.818 & 0.021 & 0.012 & \best{0.004} & 0.012 & 0.011 & 0.015 & 0.032 \\
24 ($4\nu_{11}$) & 1453.394 & 0.024 & 0.014 & \best{0.004} & 0.013 & 0.011 & 0.015 & 0.036 \\
25 ($\nu_{7}$) & 1483.219 & 0.006 & 0.008 & 0.004 & 0.012 & -0.005 & \best{-0.001} & -0.047 \\
26 ($\nu_{7}$) & 1483.220 & 0.007 & 0.033 & 0.003 & 0.042 & \best{0.001} & 0.002 & 0.010 \\
27 ($\nu_{4} + 2\nu_{11}$) & 1620.199 & 0.023 & 0.011 & \best{0.003} & 0.011 & 0.020 & 0.015 & 0.036 \\
28 ($\nu_{4} + 2\nu_{11}$) & 1620.199 & 0.024 & 0.011 & \best{0.004} & 0.012 & 0.021 & 0.015 & 0.037 \\
29 ($\nu_{4} + 2\nu_{11}$) & 1620.744 & 0.044 & 0.026 & \best{0.002} & 0.028 & 0.032 & 0.034 & 0.059 \\
30 ($\nu_{3} + \nu_{11}$) & 1749.519 & 0.103 & \best{0.040} & 0.099 & 0.058 & 0.048 & 0.060 & 0.101 \\
31 ($\nu_{3} + \nu_{11}$) & 1749.519 & 0.106 & 0.084 & 0.111 & 0.067 & 0.064 & \best{0.062} & 0.123 \\
32 ($\nu_{9} + 2\nu_{11}$) & 1756.412 & 0.039 & 0.038 & 0.157 & 0.039 & 0.034 & \best{0.015} & 0.042 \\
33 ($\nu_{9} + 2\nu_{11}$) & 1756.412 & 0.040 & 0.041 & 0.174 & 0.041 & 0.039 & \best{0.015} & 0.043 \\
34 ($\nu_{9} + 2\nu_{11}$) & 1757.119 & 0.047 & 0.034 & 0.110 & 0.034 & 0.024 & \best{0.021} & 0.051 \\
35 ($\nu_{9} + 2\nu_{11}$) & 1757.119 & 0.048 & 0.040 & 0.112 & 0.043 & 0.034 & \best{0.022} & 0.052 \\
36 ($\nu_{9} + 2\nu_{11}$) & 1759.760 & 0.121 & 0.078 & 0.127 & 0.077 & \best{0.060} & 0.073 & 0.144 \\
37 ($\nu_{9} + 2\nu_{11}$) & 1759.760 & 0.124 & 0.080 & 0.131 & 0.079 & \best{0.061} & 0.080 & 0.150 \\
38 ($2\nu_{4}$) & 1785.177 & -0.041 & -0.052 & -0.057 & -0.051 & -0.039 & -0.048 & \best{-0.030} \\
39 ($5\nu_{11}$) & 1816.786 & 0.006 & 0.005 & 0.006 & 0.005 & \best{0.004} & \best{0.004} & 0.016 \\
40 ($5\nu_{11}$) & 1816.786 & 0.006 & 0.005 & 0.006 & 0.005 & \best{0.004} & \best{0.004} & 0.016 \\
41 ($5\nu_{11}$) & 1818.938 & 0.030 & 0.019 & \best{0.005} & 0.019 & 0.017 & 0.021 & 0.042 \\
42 ($5\nu_{11}$) & 1818.939 & 0.029 & 0.018 & \best{0.007} & 0.018 & 0.016 & 0.020 & 0.043 \\
43 ($5\nu_{11}$) & 1820.016 & 0.033 & 0.019 & \best{0.005} & 0.020 & 0.018 & 0.021 & 0.050 \\
44 ($5\nu_{11}$) & 1820.016 & 0.034 & 0.022 & \best{0.005} & 0.022 & 0.018 & 0.024 & 0.051 \\
45 ($\nu_{7} + \nu_{11}$) & 1844.245 & 0.079 & 0.076 & 0.096 & 0.078 & \best{0.010} & 0.045 & 0.086 \\
46 ($\nu_{7} + \nu_{11}$) & 1844.316 & 0.092 & 0.099 & 0.052 & 0.093 & \best{0.012} & 0.059 & 0.084 \\
47 ($\nu_{7} + \nu_{11}$) & 1844.317 & 0.093 & 0.100 & 0.098 & 0.093 & \best{0.014} & 0.071 & 0.141 \\
48 ($\nu_{7} + \nu_{11}$) & 1844.676 & 0.080 & 0.108 & 0.094 & 0.080 & \best{0.013} & 0.046 & -0.048 \\
49 ($\nu_{4} + \nu_{9}$) & 1931.514 & 0.023 & \best{0.001} & 0.008 & 0.009 & 0.016 & 0.019 & 0.035 \\
50 ($\nu_{4} + \nu_{9}$) & 1931.514 & 0.024 & 0.015 & \best{0.009} & 0.026 & 0.017 & 0.026 & 0.039 \\
51 ($\nu_{4} + 3\nu_{11}$) & 1981.815 & 0.035 & 0.022 & \best{0.006} & 0.022 & 0.028 & 0.024 & 0.051 \\
52 ($\nu_{4} + 3\nu_{11}$) & 1981.815 & 0.036 & 0.023 & \best{0.007} & 0.023 & 0.029 & 0.025 & 0.053 \\
53 ($\nu_{4} + 3\nu_{11}$) & 1982.818 & 0.076 & 0.048 & \best{0.003} & 0.049 & 0.052 & 0.061 & 0.090 \\
54 ($\nu_{4} + 3\nu_{11}$) & 1982.818 & 0.078 & 0.048 & \best{0.003} & 0.050 & 0.053 & 0.061 & 0.090 \\
55 ($2\nu_{9}$) & 2057.044 & 0.019 & \best{0.015} & \best{0.015} & 0.016 & \best{0.015} & 0.016 & 0.036 \\
56 ($2\nu_{9}$) & 2065.265 & 0.011 & 0.010 & 0.009 & 0.011 & \best{0.008} & \best{0.008} & 0.025 \\
57 ($2\nu_{9}$) & 2065.265 & 0.015 & 0.013 & 0.038 & 0.018 & 0.011 & \best{0.009} & 0.027 \\
58 ($\nu_{3} + 2\nu_{11}$) & 2111.364 & 0.171 & 0.129 & 0.173 & 0.125 & 0.101 & \best{0.097} & 0.151 \\
59 ($\nu_{3} + 2\nu_{11}$) & 2111.364 & 0.189 & 0.130 & 0.174 & 0.126 & 0.111 & \best{0.097} & 0.162 \\
60 ($\nu_{3} + 2\nu_{11}$) & 2112.281 & 0.215 & 0.164 & 0.151 & 0.168 & \best{0.116} & 0.141 & 0.215 \\
61 ($\nu_{9} + 3\nu_{11}$) & 2119.307 & 0.051 & 0.051 & 0.209 & 0.052 & 0.046 & \best{0.022} & 0.055 \\
62 ($\nu_{9} + 3\nu_{11}$) & 2119.307 & 0.052 & 0.053 & 0.210 & 0.055 & 0.050 & \best{0.023} & 0.056 \\
63 ($\nu_{9} + 3\nu_{11}$) & 2120.521 & 0.122 & 0.126 & 0.187 & 0.121 & 0.111 & \best{0.074} & 0.118 \\
64 ($\nu_{9} + 3\nu_{11}$) & 2120.521 & 0.132 & 0.130 & 0.199 & 0.134 & 0.125 & \best{0.082} & 0.124 \\
65 ($\nu_{9} + 3\nu_{11}$) & 2120.889 & 0.079 & 0.063 & 0.144 & 0.069 & 0.066 & \best{0.030} & 0.065 \\
66 ($\nu_{9} + 3\nu_{11}$) & 2122.816 & 0.178 & 0.124 & 0.170 & 0.125 & \best{0.091} & \best{0.091} & 0.157 \\
67 ($\nu_{9} + 3\nu_{11}$) & 2122.816 & 0.184 & 0.124 & 0.179 & 0.159 & \best{0.094} & 0.095 & 0.177 \\
68 ($\nu_{9} + 3\nu_{11}$) & 2123.282 & 0.249 & 0.194 & 0.185 & 0.198 & \best{0.150} & 0.151 & 0.255 \\
69 ($2\nu_{4} + \nu_{11}$) & 2142.444 & -0.032 & -0.052 & -0.063 & -0.050 & -0.035 & -0.044 & \best{-0.021} \\
70 ($2\nu_{4} + \nu_{11}$) & 2142.444 & -0.031 & -0.050 & -0.037 & -0.048 & -0.032 & -0.043 & \best{-0.012} \\
71 ($6\nu_{11}$) & 2183.617 & 0.008 & 0.007 & 0.009 & 0.007 & \best{0.006} & \best{0.006} & 0.019 \\
72 ($6\nu_{11}$) & 2183.617 & 0.008 & 0.007 & 0.009 & 0.007 & \best{0.006} & \best{0.006} & 0.020 \\
73 ($6\nu_{11}$) & 2186.117 & 0.041 & 0.026 & \best{0.008} & 0.027 & 0.023 & 0.029 & 0.052 \\
74 ($6\nu_{11}$) & 2186.117 & 0.041 & 0.026 & \best{0.008} & 0.027 & 0.023 & 0.029 & 0.053 \\
75 ($6\nu_{11}$) & 2187.618 & 0.054 & 0.035 & \best{0.010} & 0.036 & 0.030 & 0.039 & 0.069 \\
76 ($6\nu_{11}$) & 2187.618 & 0.055 & 0.038 & \best{0.010} & 0.038 & 0.031 & 0.041 & 0.072 \\
77 ($6\nu_{11}$) & 2188.119 & 0.057 & 0.040 & \best{0.010} & 0.042 & 0.035 & 0.043 & 0.078 \\
78 ($\nu_{7} + 2\nu_{11}$) & 2206.608 & 0.128 & 0.116 & 0.100 & 0.112 & \best{0.043} & 0.134 & 0.115 \\
79 ($\nu_{7} + 2\nu_{11}$) & 2206.615 & 0.128 & 0.130 & 0.146 & 0.108 & \best{0.084} & 0.130 & 0.115 \\
80 ($\nu_{7} + 2\nu_{11}$) & 2206.757 & -0.010 & -0.006 & 0.057 & -0.013 & -0.050 & \best{-0.002} & \best{-0.002} \\
81 ($\nu_{7} + 2\nu_{11}$) & 2206.758 & 0.008 & \best{-0.002} & 0.137 & 0.016 & 0.026 & 0.049 & 0.027 \\
82 ($\nu_{7} + 2\nu_{11}$) & 2207.541 & 0.019 & 0.073 & 0.058 & 0.035 & \best{0.008} & -0.009 & 0.064 \\
83 ($\nu_{7} + 2\nu_{11}$) & 2207.541 & 0.088 & 0.094 & 0.067 & 0.075 & \best{0.015} & 0.063 & 0.105
 \\
\end{longtable}

\section{Conclusions}
\label{sec:conclusion}

In this work, we developed and benchmarked a generalized tree tensor network framework for vibrational spectroscopy in which both the Hamiltonian and the wavefunction are represented as trees, i.e., as a TTNO and a TTNS, respectively. Within this framework, we implemented and combined block LOBPCG and block inverse iteration methods. This approach achieves sub-wavenumber accuracy for two representative benchmark systems --- a 64-dimensional bilinearly coupled oscillator model and acetonitrile as a real molecular system  --- demonstrating the capability to solve chemical physics problems with strong coupling and near-degeneracies.

For the coupled-oscillator model, T tree systematically improves on MPS at the same bond dimension, which is consistent with the all-to-all coupling pattern of the Hamiltonian. For acetonitrile, inverse iteration reduces the mean errors from the 0.298--0.436~cm$^{-1}$ range of the initial LOBPCG results to 0.026--0.054~cm$^{-1}$ across all tested topologies, and all 84 computed states end below 1~cm$^{-1}$ error. Among the tested trees, the fork-4 topology provides the most favorable overall compromise between accuracy and runtime. The results therefore show that increased tree connectivity is clearly beneficial, but that moderate branching is more effective here than simply adding as many auxiliary tensors as possible.

An important aspect of this work is the systematic comparison of tensor network topologies within a common TTNO/TTNS implementation. This enables us to examine the effect of topology in a controlled setting, while minimizing the influence of differences in code or algorithmic details. The results suggest that the choice of topology is an important practical factor in vibrational tensor-network calculations.

All implementations used in this work are available as open-source code in the PyTreeNet ecosystem. The repository containing the TTNO/TTNS eigensolvers, input files for the oscillator and acetonitrile benchmarks, and scripts to reproduce the figures is publicly available at \url{https://github.com/sunshuo987/vibcal}. This enables full reproducibility and provides a platform for further developments, such as automated tree-structure optimization, improved preconditioners, and the computation of transition intensities for IR and Raman spectra.

\begin{acknowledgments}
Shuo Sun acknowledges funding by the BMW Group.

The research is part of the Munich Quantum Valley, which is supported by the Bavarian state government with funds from the Hightech Agenda Bayern Plus.

We also like to thank the Munich Center for Quantum Science and Technology (MCQST) for support.
\end{acknowledgments}

\section*{Author Declarations}
\subsection*{Conflict of Interest Statement}
The authors have no conflicts to disclose.

\subsection*{Author Contributions}
S. Sun: Conceptualization (lead); Investigation (lead); Methodology (lead); Software (lead); Validation (lead); Writing – original draft (lead). R. M. Milbradt: Conceptualization (equal); Software (lead);  Investigation (equal); Methodology (equal); Writing – original draft (equal). S. Knecht: Conceptualization (equal); Methodology (equal); Supervision (equal); Writing – review \& editing (equal). C. Kumar: Conceptualization (equal); Funding acquisition (equal); Supervision (equal); Writing - review \& editing (equal). C. B. Mendl: Conceptualization (equal); Funding acquisition (equal); Supervision (lead); Writing - review \& editing (equal).

\section*{Data Availability Statement}
The data that supports the findings of this study are openly available at \url{https://github.com/sunshuo987/vibcal}.

\appendix
\section{Scaling of Tree Tensor Networks Methods}
\label{sec: scaling}
This appendix lists the tensor scaling of the TTNS-TTNO application methods mentioned. We consider $D_O$ to be the maximum bond dimension in the TTNO, $D_S$ the maximum bond dimension of the TTNS, $\chi$ the desired bond dimension of the TTNS after the application, and $d$ the maximum physical dimension. Note that $\kappa$ is the maximum number of neighbours a node can have in a given tree. Thus $\kappa=2$ for MPS and $\kappa=3$ for T3NS. The scalings are listed in \autoref{tab:scalings}.

\begin{table}[ht]
\caption{The scaling for the different methods and different tree structures.}
\label{tab:scalings}
\begin{tabular}{c|c|c|c}
        Tree & General & MPS & T3NS \\
        \hline
        SRC & $\begin{aligned}\sum_{j=1}^{\kappa-1} & dD_S^{\kappa-j+1}D_O^{j}\chi^j + d^2D_SD_O^\kappa \chi^{\kappa-1}\\ +& d^2 D_SD_O\chi^{2\kappa-2} \end{aligned}$ & $\begin{aligned}&dD_S^2D_O\chi + d^2D_SD_O^2\chi \\ &+ d^2D_SD_O\chi^2 \end{aligned}$ & $\begin{aligned} & dD_S^3D_O\chi + dD_S^2D_O^2\chi^2 \\ & + d^2D_SD_O\chi^4 \end{aligned}$\\
        \hline
        Direct & & &\\
        Contraction & $ d^2D_S^\kappa D_O^\kappa + d^2 D_SD_O\chi^{2\kappa-2}$ & $ d^2D_S^2D_O^2 + d^2\chi^2 D_SD_O $ & $\begin{aligned} d^2D_S^3D_O^3 + d^2\chi^4D_SD_O\end{aligned}$\\
        \hline
        Variational fitting & $\begin{aligned}\sum_{j=1}^{\kappa} d^2 D_S^{2(\kappa-j+1)}D_O^{j}+d^2 D_S^{2\kappa}\end{aligned}$ & $d^2 D_S^4 D_O + d^2 D_S^{4}$ & $d^2 D_S^6 D_O + d^2 D_S^{6}$\\
        \hline 
        ALS & $\begin{aligned}\sum_{j=1}^{\kappa} d^2 D_S^{2(\kappa-j+1)}D_O^{j}+d^2 D_S^{2\kappa}\end{aligned}$ & $d^2 D_S^4 D_O + d^2 D_S^{4}$ & $d^2 D_S^6 D_O + d^2 D_S^{6}$\\
        \hline 
\end{tabular}
\end{table}

In tree topologies where physical nodes can appear on internal (non-leaf) nodes---such as full T3NS and fork-$n$ trees---each such node carries both a physical leg (of dimension $d$) and up to two virtual legs (of dimension $D_s$), resulting in tensors of size $d \times D_s^2$. In contrast, leaf-only trees place physical legs exclusively on leaf nodes, each connected to only one virtual leg, leading to tensors of size $d \times D_s$. The internal branching nodes then contribute tensors of size up to $D_s^3$. Moreover, leaf-only trees generally require a substantially larger number of internal branching nodes, often comparable to the number of physical nodes, whereas full T3NS and fork-$n$ trees require significantly fewer such nodes. Consequently, for systems requiring large local basis sizes $d$, the computational cost of full T3NS and fork-$n$ trees increases more rapidly than that of leaf-only trees. On the other hand, when $d$ remains moderate, the larger number of internal branching nodes can become a limiting factor for the efficiency of leaf-only trees.
\section{Tree structure used in the vibrational spectra calculations}
\label{sec:appendix_tree_structure}
\begin{figure}[htbp]
  \centering
  \subfloat[MPS structure\label{fig:mps_64d}]{%
  \raisebox{1.0\height}{%
\begin{tikzpicture}[
  circ/.style={circle, draw, fill=cyan!70, ultra thick, minimum size=8mm, inner sep=0pt},
  node distance=4mm                         
]
\node[circ] (c1) {1};
\node[circ, right=of c1] (c2) {2};
\node[circ, right=of c2] (c3) {3};

\node[right=2mm of c3] (dots) {$\cdots$};

\node[circ, right=2mm of dots] (c4) {62};
\node[circ, right=of c4] (c5) {63};
\node[circ, right=of c5] (c6) {64};

\draw[very thick] (c1) -- (c2) (c2) -- (c3) (c4) -- (c5) (c5) -- (c6);
\draw[very thick] (c3.east) -- ++(2mm,0);
\draw[very thick] (c4.west) -- ++(-2mm,0);

\foreach \n in {c1,c2,c3,c4,c5,c6} {
  \draw[very thick] (\n.south) -- ++(0,-3mm);
}
\end{tikzpicture}
 }}\hfill
  \subfloat[T tree structure\label{fig:t3ns64}]{%
\begin{tikzpicture}[
  circ/.style={circle, draw, fill=cyan!70, ultra thick, minimum size=8mm, inner sep=0pt},
  node distance=4mm
]
\node[circ] (L) {1};

\node (dotsL) [right=4mm of L] {$\cdots$};

\node[circ] (A) [right=4mm of dotsL] {22};
\node[circ,fill=black!30] (B) [right=of A] {};
\node[circ] (C) [right=of B] {23};

\node (dotsR) [right=4mm of C] {$\cdots$};
\node[circ] (R) [right=3.75mm of dotsR] {44};

\node[circ] (Bdown1) [below=4mm of B] {45};
\node (vdots) [below=2.5mm of Bdown1] {$\vdots$};
\node[circ] (Bdown2) [below=3.5mm of vdots] {64};

\draw[very thick] (A) -- (B) -- (C);
\draw[very thick] (B) -- (Bdown1);
\draw[very thick] (L.east) -- ++(4mm,0);  
\draw[very thick] (A.west) -- ++(-4mm,0);
\draw[very thick] (C.east) -- ++(4mm,0);  
\draw[very thick] (R.west) -- ++(-4mm,0);
\draw[very thick] (Bdown1.south) -- ++(0,-4mm);
\draw[very thick] (Bdown2.north) -- ++(0,4mm);

\foreach \n in {L, A, C, R, Bdown1, Bdown2}{
    \draw[very thick] (\n.north east) -- ++(1.5mm,1.5mm);
}
\end{tikzpicture}
}
\caption{Tensor network structures used for the 64-dimensional coupled harmonic oscillator calculations: (a) MPS, and (b) T tree structure. The blue circles denote nodes carrying physical legs, while the gray circles represent auxiliary tensor nodes.}
\label{fig:ttn_structures_64}
\end{figure}
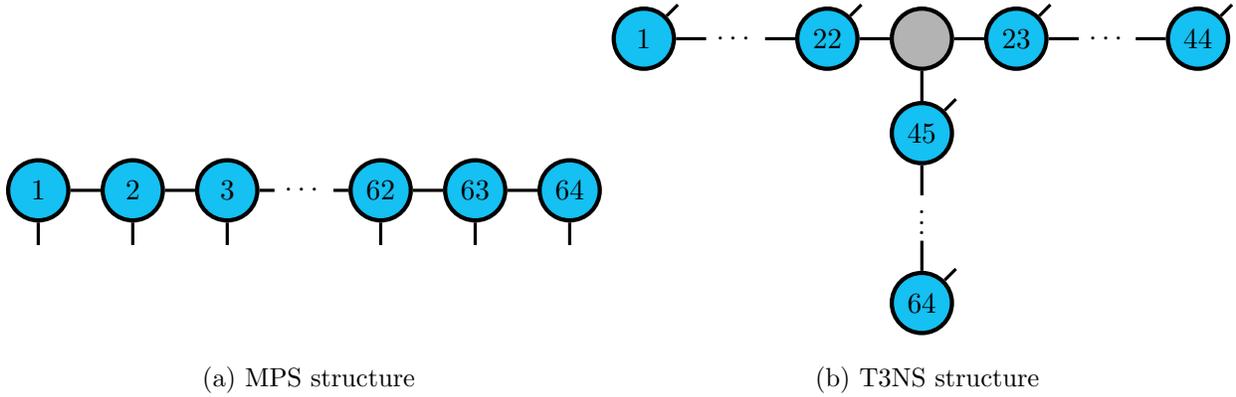

\begin{figure}[ht]
  \centering
  \subfloat[MPS structure\label{fig:mps_ch3cn}]{%
\begin{tikzpicture}[
  circ/.style={circle, draw, fill=cyan!70, ultra thick, minimum size=8mm, inner sep=0pt},
  node distance=4mm                         
]
\node[circ] (c1) {1};
\node[circ, right=of c1] (c2) {2};
\node[circ, right=of c2] (c3) {3};
\node[circ, right=of c3] (c4) {4};
\node[circ, right=of c4] (c5) {5};
\node[circ, right=of c5] (c6) {6};
\node[circ, right=of c6] (c7) {7};
\node[circ, right=of c7] (c8) {8};
\node[circ, right=of c8] (c9) {9};
\node[circ, right=of c9] (c10) {10};
\node[circ, right=of c10] (c11) {11};
\node[circ, right=of c11] (c12) {12};

\draw[very thick] (c1) -- (c2) (c2) -- (c3) (c3) -- (c4) (c4) -- (c5) (c5) -- (c6) (c6) -- (c7) (c7) -- (c8) (c8) -- (c9) (c9) -- (c10) (c10) -- (c11) (c11) -- (c12);
\draw[very thick] (c3.east) -- ++(2mm,0);
\draw[very thick] (c4.west) -- ++(-2mm,0);

\foreach \n in {c1,c2,c3,c4,c5,c6,c7,c8,c9,c10,c11,c12} {
  \draw[very thick] (\n.south) -- ++(0,-3mm);
}
\end{tikzpicture}
 }\hfill
  \subfloat[T Tree structure\label{fig:t3ns4ch3cn}]{%
\begin{tikzpicture}[
  circ/.style={circle, draw, fill=cyan!70, ultra thick, minimum size=8mm, inner sep=0pt},
  node distance=4mm
]
\node[circ] (c1) {1};
\node[circ, right=of c1] (c2) {2};
\node[circ, right=of c2] (c3) {3};
\node[circ, right=of c3] (c4) {4};
\node[circ,fill=black!30] (B) [right=of c4] {};
\node[circ] (c5) [right=of B] {5};
\node[circ, right=of c5] (c6) {6};
\node[circ, right=of c6] (c7) {7};
\node[circ, right=of c7] (c8) {8};
\node[circ] (c9) [below=4mm of B] {9};
\node[circ, below=of c9] (c10) {10};
\node[circ, below=of c10] (c11) {11};
\node[circ, below=of c11] (c12) {12};

\draw[very thick] (c1) -- (c2) (c2) -- (c3) (c3) -- (c4) (c4) --(B) (B)--(c5) (c5) -- (c6) (c6) -- (c7) (c7) -- (c8) (B) -- (c9) (c9) -- (c10) (c10) -- (c11) (c11) -- (c12);

\foreach \n in {c1,c2,c3,c4,c5,c6,c7,c8,c9,c10,c11,c12}{
    \draw[very thick] (\n.north east) -- ++(1.5mm,1.5mm);
}
\end{tikzpicture}
}

  \subfloat[Leaf-only structure ~\label{fig:leaf4ch3cn}]{%
\begin{tikzpicture}[
  circ/.style={circle, draw, fill=cyan!70, ultra thick, minimum size=8mm, inner sep=0pt},
  circa/.style={circle, draw, fill=black!30, ultra thick, minimum size=8mm, inner sep=0pt},
  node distance=4mm
]
\def\nodedist{2}
\def\physdist{0.5}
    
\node[circ,fill=black!30] (root) {};
 \foreach \j [evaluate=\j as \maxi using int(2^\j)] in {0,1,2} {%
    \foreach \i in {1,...,\maxi} {%
      \node[circa] (R\i\j) at ($(root) + (-8/\maxi*0.5+ 8/\maxi* \i,-.6*\j*\nodedist-.5*\nodedist)$){};
      \node[circa] (L\i\j) at ($(root) + (8/\maxi*0.5- 8/\maxi* \i,-.6*\j*\nodedist-.5*\nodedist)$){};
    }%
  }%
  \node[circ] (1) at ($(root) + (8/4*0.5- 8/4* 3,-.6*2*\nodedist-.5*\nodedist)$) {1};
  \node[circ] (3) at ($(root) + (-8/4*0.5+ 8/4* 2,-.6*2*\nodedist-.5*\nodedist)$) {3};
  \node[circ] (11) at ($(root) + (-8/4*0.5+ 8/4* 3,-.6*2*\nodedist-.5*\nodedist)$) {11};
  \node[circ] (12) at ($(root) + (-8/4*0.5+ 8/4* 4,-.6*2*\nodedist-.5*\nodedist)$) {12};
  \node[circ] (5) at ($(root) + (8/8*0.5- 8/8* 8,-.6*3*\nodedist-.5*\nodedist)$) {5};
  \node[circ] (6) at ($(root) + (8/8*0.5- 8/8* 7,-.6*3*\nodedist-.5*\nodedist)$) {6};
  \node[circ] (7) at ($(root) + (8/8*0.5- 8/8* 4,-.6*3*\nodedist-.5*\nodedist)$) {7};
  \node[circ] (8) at ($(root) + (8/8*0.5- 8/8* 3,-.6*3*\nodedist-.5*\nodedist)$) {8};
  \node[circ] (9) at ($(root) + (8/8*0.5- 8/8* 2,-.6*3*\nodedist-.5*\nodedist)$) {9};
  \node[circ] (10) at ($(root) + (8/8*0.5- 8/8* 1,-.6*3*\nodedist-.5*\nodedist)$) {10};
  \node[circ] (2) at ($(root) + (-8/8*0.5+ 8/8* 1,-.6*3*\nodedist-.5*\nodedist)$) {2};
  \node[circ] (4) at ($(root) + (-8/8*0.5+ 8/8* 2,-.6*3*\nodedist-.5*\nodedist)$) {4};
  \draw [very thick] (root)--(L10)--(L21)--(L42)--(5) (L42)--(6) (L21)--(1);
  \draw [very thick] (L10)--(L11)--(L22)--(7) (L22)--(8) (L11)--(L12)--(9) (L12)--(10);
  \draw [very thick] (root)--(R10)--(R21)--(11) (R21)--(12) (R11)--(3);
  \draw [very thick] (R10)--(R11)--(R12)--(2) (R12)--(4);
  \foreach \i in {1,2,3,4,5,6,7,8,9,10,11,12}{
  \draw[very thick] (\i.south) -- ++(0,-4mm);
  }
\end{tikzpicture}
}
\caption{Tensor network structures used for the acetonitrile ($\mathrm{CH_3CN}$) vibrational spectra calculations:
(a) MPS, (b) T tree tensor network, and (c) leaf-only tree.
The blue circles denote nodes carrying physical legs, while the gray circles represent auxiliary tensor nodes.
Panel (c) was generated by the authors based on the leaf-only tree structure reported in Ref.~\onlinecite{Larsson2019}.}
\label{fig:ttn_structures_ch3cn_1}
\end{figure}

\begin{figure}[ht]
  \centering
  \subfloat[Fork-3 structure~\label{fig:fork3_ch3cn}]{%
\begin{tikzpicture}[
  circ/.style={circle, draw, fill=cyan!70, ultra thick, minimum size=8mm, inner sep=0pt},
  node distance=4mm                         
]
\node[circ,fill=black!30] (b1) {};
\node[circ,fill=black!30, right=of b1] (b2){};
\node[circ,fill=black!30, right=of b2] (b3){};
\node[fill=white, right=of b3] (b4){};
\node[circ, below=of b1] (c1) {1};
\node[circ, below=of c1] (c2) {2};
\node[circ, below=of c2] (c3) {3};
\node[circ, below=of c3] (c4) {4};
\node[circ, below=of b2] (c5) {5};
\node[circ, below=of c5] (c6) {6};
\node[circ, below=of c6] (c7) {7};
\node[circ, below=of c7] (c8) {8};
\node[circ, below=of b3] (c9) {9};
\node[circ, below=of c9] (c10) {10};
\node[circ, below=of c10] (c11) {11};
\node[circ, below=of c11] (c12) {12};

\draw[very thick] (b1) -- (b2) (b2) -- (b3) (b1) -- (c1) (c1) -- (c2) (c2) -- (c3) (c3) -- (c4) (b2) -- (c5) (c5) -- (c6) (c6) -- (c7) (c7) -- (c8) (b3) -- (c9) (c9) -- (c10) (c10) -- (c11) (c11) -- (c12);
\foreach \n in {c1,c2,c3,c4,c5,c6,c7,c8,c9,c10,c11,c12}{
    \draw[very thick] (\n.north east) -- ++(1.5mm,1.5mm);
}
\end{tikzpicture}}
\hfill
\subfloat[Fork-4 structure\label{fig:fork4_ch3cn}]{
\begin{tikzpicture}[
  circ/.style={circle, draw, fill=cyan!70, ultra thick, minimum size=8mm, inner sep=0pt},
  node distance=4mm                         
]
\node[circ,fill=black!30] (b1) {};
\node[circ,fill=black!30, right=of b1] (b2){};
\node[circ,fill=black!30, right=of b2] (b3){};
\node[circ,fill=black!30, right=of b3] (b4){};
\node[circ, below=of b1] (c1) {1};
\node[circ, below=of c1] (c2) {2};
\node[circ, below=of c2] (c3) {3};
\node[circ, below=of b2] (c4) {4};
\node[circ, below=of c4] (c5) {5};
\node[circ, below=of c5] (c6) {6};
\node[circ, below=of b3] (c7) {7};
\node[circ, below=of c7] (c8) {8};
\node[circ, below=of c8] (c9) {9};
\node[circ, below=of b4] (c10) {10};
\node[circ, below=of c10] (c11) {11};
\node[circ, below=of c11] (c12) {12};

\draw[very thick] (b1) -- (b2) (b2) -- (b3) (b3) -- (b4) (b1) -- (c1) (c1) -- (c2) (c2) -- (c3) (b2) -- (c4) (c4) -- (c5) (c5) -- (c6) (b3) -- (c7) (c7) -- (c8) (c8) -- (c9) (b4) -- (c10) (c10) -- (c11) (c11) -- (c12);
\foreach \n in {c1,c2,c3,c4,c5,c6,c7,c8,c9,c10,c11,c12}{
    \draw[very thick] (\n.north east) -- ++(1.5mm,1.5mm);
}
\end{tikzpicture}}
\hfill
\subfloat[Fork-6 structure\label{fig:fork6_ch3cn}]{%
\begin{tikzpicture}[
  circ/.style={circle, draw, fill=cyan!70, ultra thick, minimum size=8mm, inner sep=0pt},
  node distance=4mm                         
]
\node[circ,fill=black!30] (b1) {};
\node[circ,fill=black!30, right=of b1] (b2){};
\node[circ,fill=black!30, right=of b2] (b3){};
\node[circ,fill=black!30, right=of b3] (b4){};
\node[circ,fill=black!30, right=of b4] (b5){};
\node[circ,fill=black!30, right=of b5] (b6){};
\node[circ, below=of b1] (c1) {1};
\node[circ, below=of c1] (c2) {2};
\node[circ, below=of b2] (c3) {3};
\node[circ, below=of c3] (c4) {4};
\node[circ, below=of b3] (c5) {5};
\node[circ, below=of c5] (c6) {6};
\node[circ, below=of b4] (c7) {7};
\node[circ, below=of c7] (c8) {8};
\node[circ, below=of b5] (c9) {9};
\node[circ, below=of c9] (c10) {10};
\node[circ, below=of b6] (c11) {11};
\node[circ, below=of c11] (c12) {12};

\draw[very thick] (b1) -- (b2) (b2) -- (b3) (b3) -- (b4) (b4) -- (b5) (b5) -- (b6) (b1) -- (c1) (c1) -- (c2) (b2) -- (c3) (c3) -- (c4) (b3) -- (c5) (c5) -- (c6) (b4) -- (c7) (c7) -- (c8) (b5) -- (c9) (c9) -- (c10) (b6) -- (c11) (c11) -- (c12);
\foreach \n in {c1,c2,c3,c4,c5,c6,c7,c8,c9,c10,c11,c12}{
    \draw[very thick] (\n.north east) -- ++(1.5mm,1.5mm);
}
\end{tikzpicture}}
\hfill
\subfloat[Full T3NS structure\label{fig:fullt3ns_ch3cn}]{%
\begin{tikzpicture}[
  circ/.style={circle, draw, fill=cyan!70, ultra thick, minimum size=8mm, inner sep=0pt},
  circa/.style={circle, draw, fill=black!30, ultra thick, minimum size=8mm, inner sep=0pt},
  node distance=4mm                       
]

\def\nodedist{2}
\def\physdist{0.5}
    
\node[circ] (c1) {1};
\node[circa] (b1) at ($(c1) + (-2.4, -1)$) {};
\node[circa] (b2) at ($(c1) + (2.4, -1)$) {};
\node[circ] (c2) at ($(b1)+ (-1.2, -1)$) {2};
\node[circ] (c5) at ($(b1)+ (1.2, -1)$) {5};
\node[circ] (c8) at ($(b2)+ (-1.2, -1)$) {8};
\node[circ] (c11) at ($(b2)+ (1.2, -1)$) {11};
\node[circa, below=of c2] (b3) {};
\node[circa, below=of c5] (b4) {};
\node[circa, below=of c8] (b5) {};
\node[circa, below=of c11] (b6) {};
\node[circ] (c3) at ($(b3) + (-0.6,-1)$) {3};
\node[circ] (c4) at ($(b3) + (0.6,-1)$) {4};
\node[circ] (c6) at ($(b4) + (-0.6,-1)$) {6};
\node[circ] (c7) at ($(b4) + (0.6,-1)$) {7};
\node[circ] (c9) at ($(b5) + (-0.6,-1)$) {9};
\node[circ] (c10) at ($(b5) + (0.6,-1)$) {10};
\node[circ] (c12) at ($(b6) + (0,-1)$) {12};

\draw[very thick] (c1) -- (b1) (c1) -- (b2) (b1) -- (c2) (b1) -- (c5) (b2) -- (c8) (b2) -- (c11) (c2)--(b3) (c5)--(b4) (c8)--(b5) (c11)--(b6) (b3) -- (c3) (b3) -- (c4) (b4) -- (c6) (b4) -- (c7) (b5) -- (c9) (b5) -- (c10) (b6) -- (c12);
\foreach \n in {c1,c2,c3,c4,c5,c6,c7,c8,c9,c10,c11,c12}{
    \draw[very thick] (\n.north east) -- ++(1.5mm,1.5mm);
}
\end{tikzpicture}}
\caption{Tensor network structures used for the acetonitrile ($\mathrm{CH_3CN}$) vibrational spectra calculations:
(a) fork-3, (b) fork-4, (c) fork-6 and (d) full T3NS tree.
The blue circles denote nodes carrying physical legs, while the gray circles represent auxiliary tensor nodes.
Panels (a)--(c) were generated by the authors and were motivated by previously reported fork-tree, also known as comb-tree, structures.~\cite{Bauernfeind2017,Chepiga2019}
Panel (d) was generated by the authors and was inspired by the T3NS tree structure reported in Ref.~\onlinecite{gunst2018t3ns}.}
\label{fig:ttn_structures_ch3cn_2}
\end{figure}
\clearpage 
\bibliography{reference}
\end{document}